\documentclass[journal]{IEEEtran}
\usepackage{subcaption}
\usepackage{multirow}
\usepackage[]{algorithm2e}
\usepackage{color}

\usepackage{cite}

\ifCLASSINFOpdf
  \usepackage[pdftex]{graphicx}
  \DeclareGraphicsExtensions{.pdf,.jpeg,.png}
\else
\fi
\usepackage[cmex10]{amsmath}
\usepackage{url} 

\usepackage{listings}
\begin{document}
%
\title{Virtual Machine Introspection Based Malware Behavior Profiling and Family Grouping}
%
%
%
\author{Shun-Wen~Hsiao,
        Yeali~S.~Sun,
        and~Meng~Chang~Chen
\thanks{Shun-Wen Hsiao is with the Department of Management Information Systems, National Chengchi University, Taiwan. e-mail: hsiaom@nccu.edu.tw}
\thanks{Meng Chang Chen is with the Institute of Information Science, Academia Sinica, Taiwan. e-mail: mcc@iis.sinica.edu.tw}
\thanks{Yeali S. Sun is with the Department of Information Management, National Taiwan University, Taiwan. email: sunny@ntu.edu.tw}
\thanks{Manuscript received May 4, 2017.}}
%
%

\markboth{arXiv, May~2017}%
{Hsiao \MakeLowercase{\textit{et al.}}: Virtual Machine Introspection Based Malware Behavior Profiling and Family Grouping}
%



\maketitle

\begin{abstract}
The proliferation of malwares have been attributed to the alternations of the original malware source codes. The malwares alternated from the same origin share some intrinsic behaviors and form a malware family. Expediently, identifying its malware family when a malware is first seen can provide useful clues to mitigating the threat. In this paper, a malware profiler (VMP) is proposed to profile the execution behaviors of a malware at the runtime by leveraging the virtual machine introspection (VMI) technique. The VMP inserts a plug-in inside the virtual machine monitor (VMM) to record the invoked Windows API calls with the parameters and return values as the profile of a malware. Based on the profiles, we then adopt a distance measurement and a phylogenetic tree construction method to discover the malware behavior groups. As expected, our study shows the malwares from a malware family are similar to each other and distinct from other malware families as well as the benign software. We then examines the goodness of the family grouping method of the VMP against existing anti-malware detection engines and some well-known grouping methods. We propose a novel peer voting method for evaluating the result of family grouping and the evaluation shows VMP is better than almost all of the compared anti-malware engines. At last, we establish a malware profiling website based on the proposed VMP for the public use.
\end{abstract}


%
\IEEEpeerreviewmaketitle

\section{Introduction}
\label{sec:intro}
%
%
%
%


\IEEEPARstart{A}{} malware (or malicious software) \cite{Egele:2012}, such as computer virus, Internet worm, trojan horse, and botnet, is developed to be planted into a target host stealthily by exploiting software vulnerability or employing social engineering plots in order to disrupt infected host operation or network service, modify or destroy software or data, steal sensitive information, or take control of the host. The loss, tangible or intangible, from the damages caused by malware, is so drastic that effective malware defense solutions are direly demanded.

The construction of malware requires intensive knowledge in the computer and network systems, as well as programming skill. Consequently, the current practices of the proliferation of malware programs are mostly from modifying existing malwares or being custom-built by the culprits who control the program codes. The malwares alternated from the same origin share some intrinsic behaviors. By knowing the origin of a newly detected malware, it helps to forge a solution to mitigate or neutralize the malware planted in the infected hosts, or even to trace back to the attack origin. We call the malwares rooted from the same origin as a malware family, and the grouping same behavior into a malware family becomes a critical issue in malware detection, defense, and forensics. The purpose of this paper is to propose a virtual machine introspection (VMI) \cite{vmi-based,Out-of-the-Box} based solution to profile the runtime behavior of malwares in terms of Windows API call sequences. Then, we investigate the behavior grouping phenomenon of the malwares by the generated profiles with a hierarchical clustering method to reveal the malware family structure. A Pairwise Classification Score (PCS) method is proposed to evaluate the goodness of different malware family classification results. Our study provide a solution includes runtime profiling, behavior grouping and classification evaluation, which is different from most of the previous works in malware detection or behavior analysis.



Once a new malware is detected, a security expert needs to answer certain critical questions, such as ``how was the malware planted into the infected host?'', ``did it access any important or private information?'', ``did it change any system configuration?'', ``is this malware similar to any known malware?'', ``can we rely on past solutions to mitigate it?'' in order to appraise possible incurred damages and curate a solution to mitigate or neutralize the malware. It is not trivial to answer the questions by examining the infected system, reviewing the system log and malware binary. It needs properly designed tools to monitor malware execution, analyze its behavior, identify its unique characteristics, cross-validate with other known malwares, and identify its malware family to assist security experts to fully explore the malware. The proposed VMP scheme is designed to achieve the above goals from profiling to analyzing.

\begin{figure}[htbp]
\centering
\includegraphics[width=0.45\textwidth]{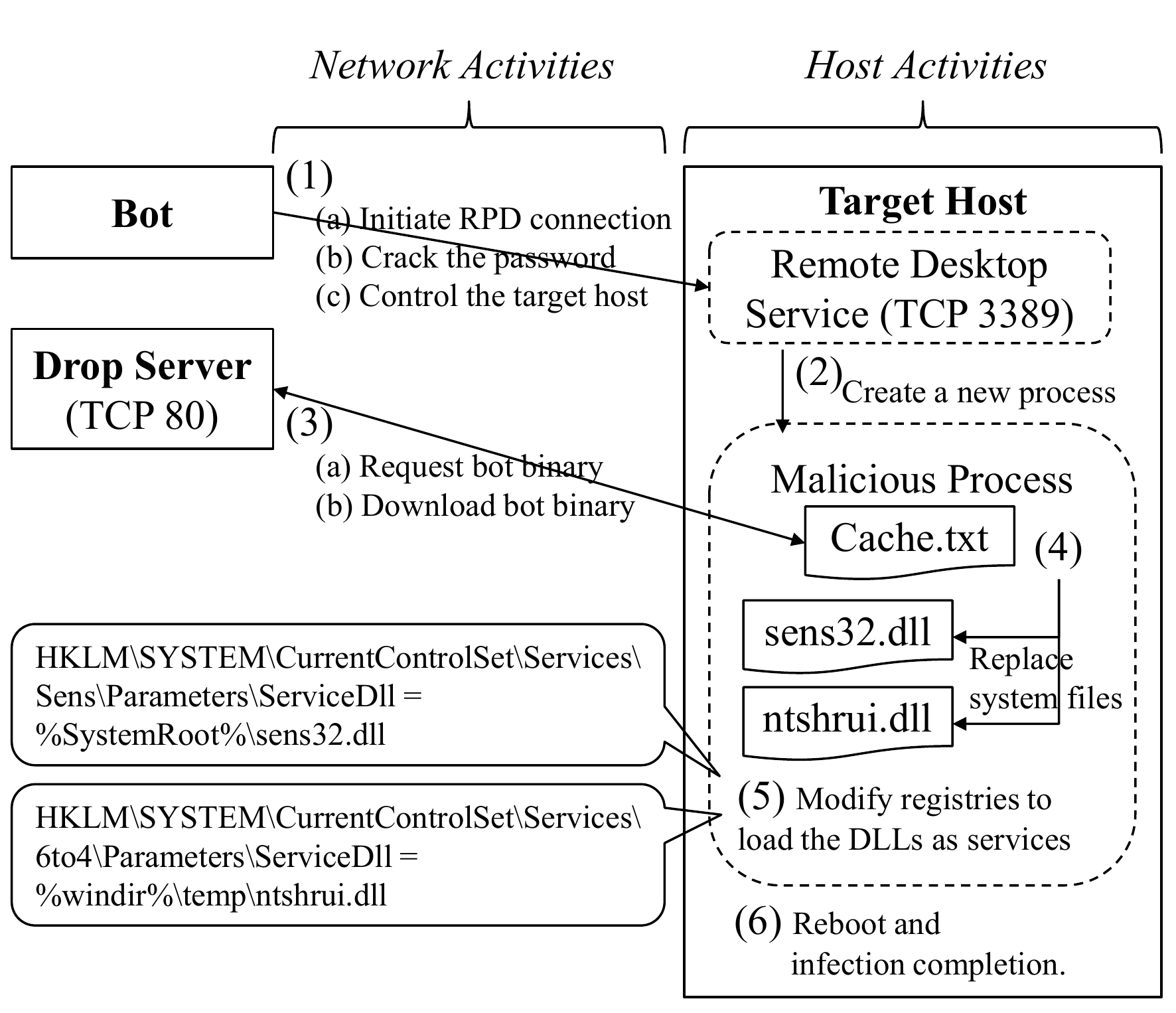}
\caption{The infection procedure of W32.Morto.A.}
\label{fig_motoa}
\end{figure}

Fig. \ref{fig_motoa} presents the behavior of a bot malware, W32.Morto.A found in late 2011, with its attack procedure and the execution trace at the infected host. The figure shows the runtime infection procedure that W32.Morto.A exploits a vulnerable host (step 1), creates and launches a temporary malicious process in the infected host (step 2), obtains malware binary from network (step 3), replaces benign system files (step 4), and makes itself as a resident service (step 5) after next system boot up (step 6). During the infection process, the bot performs message exchanges when it tries to control the host, creates a malicious process and modifies the registries to load the malicious code, which is useful for further analysis to identify the behavior of malware and its family.

In practice, there are two main streams of malware behavior analysis: static analysis and dynamic analysis \cite{Spyware,Egele:2012}. The static analysis focuses on analyzing the program code to generate static execution flow without actually executing the program and, in some case, seek for any possible logic flaws, coding errors, and vulnerable codes \cite{million,Feng}. On the other hand, the dynamic analysis \cite{Egele:2012,Gandotra} executes the target program to facilitate the collection of execution trace in various execution contexts. Although the dynamic analysis approach may not run the program in all possible contexts, it is still the only approach to obtain runtime comprehend malware behavior, such as receiving a message from Internet, loading an external library and deleting files at the runtime.

Traditionally, dynamic analysis is often performed by employing software sandbox \cite{cwsandbox} or virtual machine (VM), that both provide certain isolation between the target program and the host. Sandbox is a simulated execution environment with tight control, while a virtual machine is a hardware/software emulation of a computer system that user is able to install its own guest operating system to execute the target program as if it ran on a physical machine. 

A malware \textit{profile} is a collection of information collected during the malware execution, adequate for further analysis, and malware \textit{profiling} is the methodology and mechanism to generate a malware profile. The prerequisite of malware profiling is to have a system to be able to monitor, control and perform instructed instrumentation, and to make the execution environment as close to a real machine as possible. Consequently, the virtual machine technique is preferred in this study as it meets the above requirements. In addition, the virtual machine allows custom-made plug-ins to be embedded in the virtual machine monitor to facilitate execution trace extraction, which adds flexibility in malware profiling.

Rather than recording the low-level CPU instructions or system calls as the profiling subject, our profiler provides a higher level execution semantics --- Windows Application Programming Interface (Windows API) --- to describe a Windows malware behavior. From the past malware behavior analysis results \cite{cwsandbox,current}, malwares may use Windows APIs to access system resources, such as files, process, network information, and the registry. Hence, we hook the Windows API functions at the virtualization layer (i.e., VMM) to intercept the targeted malware at the runtime and record its invoked API calls. Moreover, we implemented a dynamic hooking mechanism that can retrieve the parameters and return value of a Windows APIs from the runtime stack, which needs to dedicate control of instrumentation and the traversal of memory.

It is common that hundred millions of CPU instructions are executed within minutes, which is overloaded and tedious for behavior analysis. The history of invoked system calls of a program is a reasonable alternative, which can be obtained from virtual machine straightforwardly \cite{Nitro}. However, with system calls alone, it still loses high-level information for security experts to comprehend the malware intents. Even with raw parameters, it still lacks semantic information for profile analysis. For instance, with system call \texttt{NtOpenFile(0x6c040000, 0x1001000, 0x74e20502, ...)}, it indicates the program opens a file from a path string pointed by \texttt{0x74e20502} with access right \texttt{0x1001000}, and gets a file handler stored in \texttt{0x6c040000}. However, reading such low-level information is not an easy work.


On the contrary, using Windows API sequences as a profile could reveal clear semantics of execution. For instance,  \texttt{LoadLibrary('SHELL32.dll')} clearly specifies the file \texttt{SHELL32.dll} to be opened and loaded is a shared library. For another example, \texttt{CreateFile('s7785.exe', GENERIC\_WRITE)} specifies the file \texttt{s7785.exe} is created for writing. Hence, the proposed VMP system is designed to obtain all such semantics associated with APIs and their parameters. (Note that in this paper, Microsoft Windows and Windows API are used as examples to exemplify the proposed design, while the design can be implemented in another context, such as Linux and C library.) 

From the past malware behavior analysis works \cite{cwsandbox,view,dynamic,temu,temu2}, malwares use APIs to access system resources, including file, process, network, Windows registry, etc. These APIs provide categorized services and libraries \cite{winapiindex} to offer convenience to programmers; security experts can grasp the semantics of APIs, and further infer the intents of malwares.


\begin{figure}[tbp]
\begin{lstlisting}[language=XML, frame=none, basicstyle=\ttfamily\small, numbers=left, breaklines=true,  showspaces=false, linewidth=0.48\textwidth, morekeywords={Profile,Meta,Execution,CreateFile,RegQueryValue,LoadLibrary,OpenProcess,Hash,Process_id,Duration,RegCreateKey,RegSetValue,DeleteFile,CreateProcessInternal,CreateProcess}]
<?xml version="1.0"?>
<Profile>
<Meta>
<Hash>61fd4cac9f5429d14d015e7632e3514a</Hash>
<Process_id>1524</Process_id>
<Duration>300</Duration>
</Meta>
<Execution>
<CreateFile hName="C:\DOCUME~1\ants\LOCALS~1\Temp\n7785\s7785.exe" desiredAccess="GENERIC_WRITE" creationDisposition="CREATE_ALWAYS" Return="SUCCESS" Time="317560000" />
<LoadLibrary lpFileName="SHELL32.dll" Return="SUCCESS" Time="339720000" />
<RegQueryValue hKey="HKCU\Software\Microsoft\Windows\ShellNoRoam\MUICache\C:\DOCUME~1\ants\LOCALS~1\Temp\n7785\s7785.exe" Return="FAILURE" Time="341100000" />
<RegSetValue hKey="HKCU\Software\Microsoft\Windows\ShellNoRoam\MUICache\C:\DOCUME~1\ants\LOCALS~1\Temp\n7785\s7785.exe" type="REG_SZ" data="install manager" Return="SUCCESS" Time="341350000" />
<CreateProcessInternal lpApplicationName="C:\DOCUME~1\ants\LOCALS~1\Temp\n7785\s7785.exe" lpCommandLine="C:\DOCUME~1\ants\LOCALS~1\Temp\n7785\s7785.exe ins.exe /e11831362 /u50d1d9d5-cf90-407c-820a-35e05bc06f2f /v" Return="SUCCESS" dwProcessId="1276" dwThreadId="1272" Time = "341600000" />
</Execution>
</Profile>
\end{lstlisting}
\caption{A (partial) malware profile example of a variant in Morstar family.}
\label{fig:xml}
\end{figure}


Fig. \ref{fig:xml} is a partial malware profile generated by the proposed VMP of a variant in the Morstar family that it shows some meta data (between line 3 to 7 that specifies the MD5 value and the process ID of the target malware, and the profiling duration is 300 seconds) and a few invoked Windows APIs with input parameters and return values, as well as the timestamps, to exemplify the format and content of malware profile. From the profile, it can be seen clearly that the Morstar variant successfully creates a file \texttt{s7785.exe} (at line 9), and loads \texttt{SHELL32.dll} (at line 10). Then it checks and sets registry value for \texttt{s7785.exe} (at lines 11 and 12), and finally creates a process to run \texttt{s7785.exe} (at line 13). Note that the spawned process will be recorded by the VMP in a separate profile as well.

Multiple malware profiles are then used for similarity analysis to group similar malwares into the associated malware behavior group (or simply `behavior family'). Then, the characteristics of each malware group can be extracted by identifying the common behaviors among its group members. In this study, the Jaccard Similarity Coefficient is adopted for similarity analysis, and a phylogenetic tree of malwares is generated to exhibit the family structure and similarities among family members. This work provides a new enabling technique to better understand the malware behavior that is anticipated to complement the existing malware detection and defense techniques.

There are several novel design principles implemented in the proposed malware profiling and analysis system. 

\begin{itemize}
\item \textbf{Taint tracking.} The design of VMP is influenced by the concept of tainted analysis \cite{temu} to track the access entities, e.g., file and registry. For example, if a process reads a file that was previously written by the targeted malware, the profiling system will also automatically profile this process.
\item \textbf{Spawned process.} The spawned processes created by malware process are tracked as well. A sophisticated malware might launch multiple processes to accomplish a job cooperatively that tracking spawned processes to give a complete view of the malware.
\item \textbf{Runtime value.} The VMP system not only records the names of invoked API but also retrieves the memory stack and CPU registers to obtain API parameters and return values.
\item \textbf{Transparency.} The profiling system is embedded in the VMM as a part of virtualization software so that the targeted malware may not aware of the existence of our profiling mechanism, and will not take defense or deceiving activities.
\item \textbf{Behavior group analysis.} The malware profile is useful for functional, structural, and evolutionary malware discovery. The phylogenetic tree composes malware behavior families and presents their relationships.
\item \textbf{Novel evaluation method for malware classification.} As there is no benchmark or golden rule for malware family grouping, a peer voting based evaluation for malware classification is proposed to appraise the grouping goodness of different anti-malware detection engines. It also solves the problem of different naming mechanisms used by different anti-virus engines, so that we can compare the classification results among different engines.
\end{itemize}

To summarize, in this work, a VMI-based Malware Profiler, VMP, is proposed to automatically record the execution trace of malware running in a virtual machine that it allows the API hooks to access high-level semantic parameter values to provide superior capability in malware family grouping and to understand malware family construction. In the experiments, with the proposed Pairwise Classification Score (PCS) method, the VMP grouping results can reveal the structure of malwares and exceed the classification results of most well-known anti-malware engines. At last, we establish a malware profiling website based on the proposed VMP for the public use.

The remainder of the paper is organized as follows. In Section \ref{sec:auto}, we present the proposed automated malware profiling scheme. Section \ref{sec:operation} is the detail design and operation of the VMP. Section \ref{sec:analysis} introduces the malware behavior analysis and behavior grouping methods. The evaluation of the behavior analysis result is in Section \ref{sec:eval}, as well as the Pairwise Classification Score result. In Section \ref{sec:related}, we review existing works of malware behavior profiling and analysis. Section \ref{sec:conclusion} contains some concluding remarks and the future works.

\section{Automated Malware Profiling}
\label{sec:auto}

\begin{figure}[!t]
\centering
\includegraphics[width=0.48\textwidth]{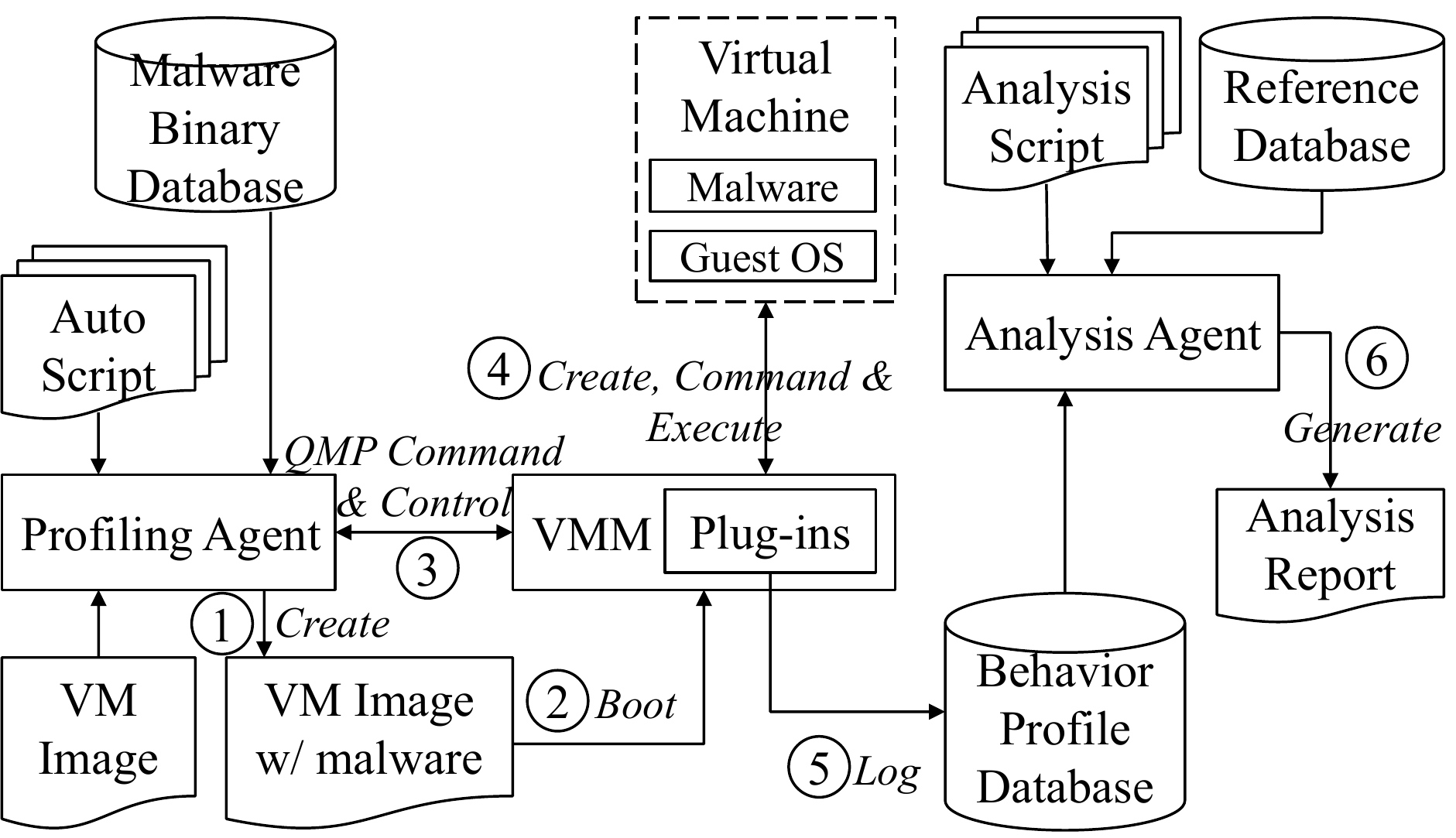}
\caption{The overview of the proposed automated malware profiling and analysis system.}
\label{fig:overview}
\end{figure}

The functional diagram of VMP is shown in Fig. \ref{fig:overview}. The automation process of malware profiling has three phases: preparation, execution, and collection. In the preparation phase, the Profiling Agent prepares a VM image for a malware to be profiled. In the execution phase, it boots the VM and executes the malware; immediately, the profiling plug-in starts to record the execution information into the profile. In the collection phase, the profile is saved in the host, and the VM is terminated and deleted. Most of the automation scripts are implemented in Python.



\subsubsection{Preparation Phase}
The Profiling Agent inserts a malware binary into a clean virtual machine image (step 1 in Fig. \ref{fig:overview}). Then, it boots the virtual machine by VMM (i.e., QEMU) with the profiling plug-ins (step 2). The Profiling Agent commands and controls the virtual machine via a specific protocol supported by the VMM (step 3). The VMM boots the guest OS in a newly created VM (step 4), where the inserted malware will be executed and the plug-ins start to record execution information for later analysis (step 5).
\subsubsection{Execution Phase}
It is not trivial to control the guest OS from the outside of a VM, especially when the Profiling Agent sometimes needs to interact with the running malware. The VMP system extends the QEMU monitor \cite{qemu} to let the Profiling Agent be able to interact with the VM and the guest OS. The QEMU Machine Protocol (QMP) \cite{qmp} is implemented in the Profiling Agent to send commands to VMM to control the execution of the VM, e.g., specifying target process, taking screen shots, turning on or off the network, moving and clicking the virtual mouse, suspending the VM, and controlling various aspects of the VM.

The Profiling Agent also issues commands via QMP to load a customized plug-in into TEMU/QEMU (Note that TEMU is an extension of QEMU \cite{temu,temu2}), which implements the callback functions for handling hooked Windows APIs. An automation script is implemented to control the virtual mouse and the keyboard attached to the VM using QMP as if a user manually interacts with the malware running in the VM. For recording network packets generated by the malware, a Wireshark client (t-shark) \cite{wireshark} is installed prior to capturing both incoming and outgoing packets.
The execution phase performs the following tasks.
\begin{itemize}
\item Load customized hooking plug-in into TEMU/QEMU.
\item Specify the targeted malware.
\item Load customized callback functions into VMM.
\item Start Wireshark client to capture packets.
\item Launch the malware by virtual mouse and keyboard controls.
\end{itemize}

\subsubsection{Collection Phase}

After execution phase, the Profiling Agent suspends the VM via QMP and collects the output, i.e., malware profile, which contains the sequence of hooked Windows API calls, a network log file having a list of network connections with source and destination IPs and network protocols, a file of the captured packets, and screen dump files of the VM. Although all the information collected above is useful for malware behavior analysis, this paper only focuses on the profile(i.e., the sequence of API calls). 

The Analysis Agent then runs the Analysis Script which contains the implementation of similarity analysis algorithm and the phylogenetic tree construction method for identifying malware behavior groups. It then writes all the results into the Analysis Report (step 6). If necessary, a script will be executed to calculate the Pairwise Classification Score for all generated profiles. Note that the Reference Database in Fig. \ref{fig:overview} is retrieved from VirusTotal.com \cite{VirusTotal} as a reference when we calculate the PCS.

\section{VMP: Virtual Machine Profiler}
\label{sec:operation}

\begin{figure}[!t]
\centering
\includegraphics[width=0.38\textwidth]{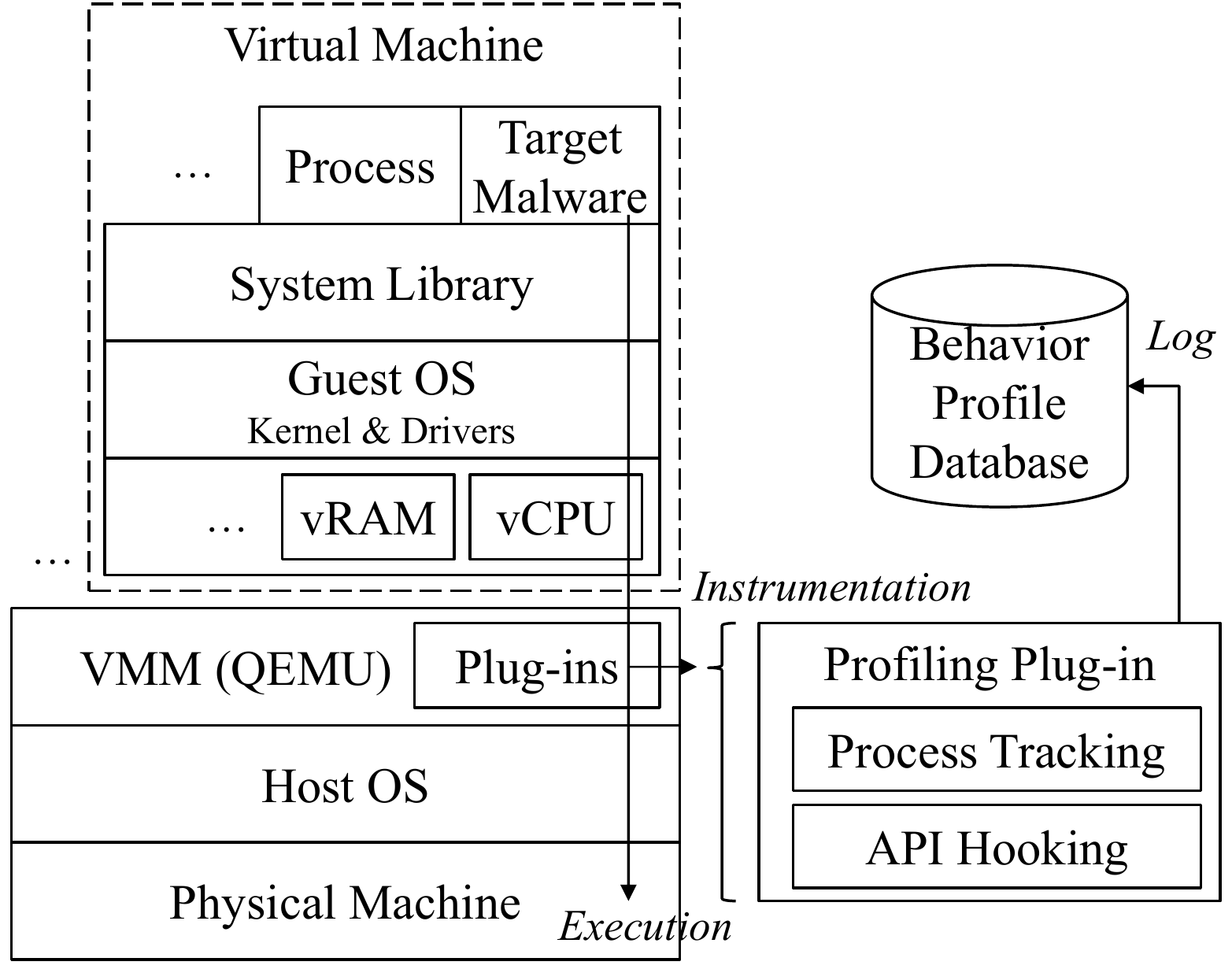}
\caption{A virtualized environment with plug-in instrumentation.}
\label{fig:vm}
\end{figure}

This section describes the design and implementation of process tracking and API hooking of the virtual machine introspection based profiler. In a virtualized environment (Fig. \ref{fig:vm}), a guest OS runs on a virtual machine created by the VMM. The VMP system is built on the top of QEMU \cite{qemu}, a free and open source to support fine-grained instrumentation at the runtime. The I/O and CPU instructions of the guest OS are all intercepted and handled by the VMM, and then are executed by the physical machine.

The profiling plug-in is implemented in the VMM to monitor the interactions between the guest OS and the VMM without modifying the OS, and also reduce the risk of being detected by the malware. Another benefit is that such design can profile all the virtual machines under the same VMM. Since the plug-ins can only read low-level VM information (i.e., CPU instruction and memory), it should bridge the semantic gap \cite{better} between the low-level data and the high-level API semantics .

The plug-in is about 3000 LOC (lines of code) written in C and C++. In addition, we modified around 1000 LOC of TEMU/QEMU to support tracking multiple processes simultaneously, which is not supported by the original version of TEMU/QEMU. In addition, for each hooked API, we implemented a customized callback function to retrieve its runtime parameters and return values, such as the name of libraries and files, the key of Windows registry, and the failure or success of API calls.

The functionalities of the plug-in are as follows.

\begin{figure}[!t]
\centering
\includegraphics[width=0.47\textwidth]{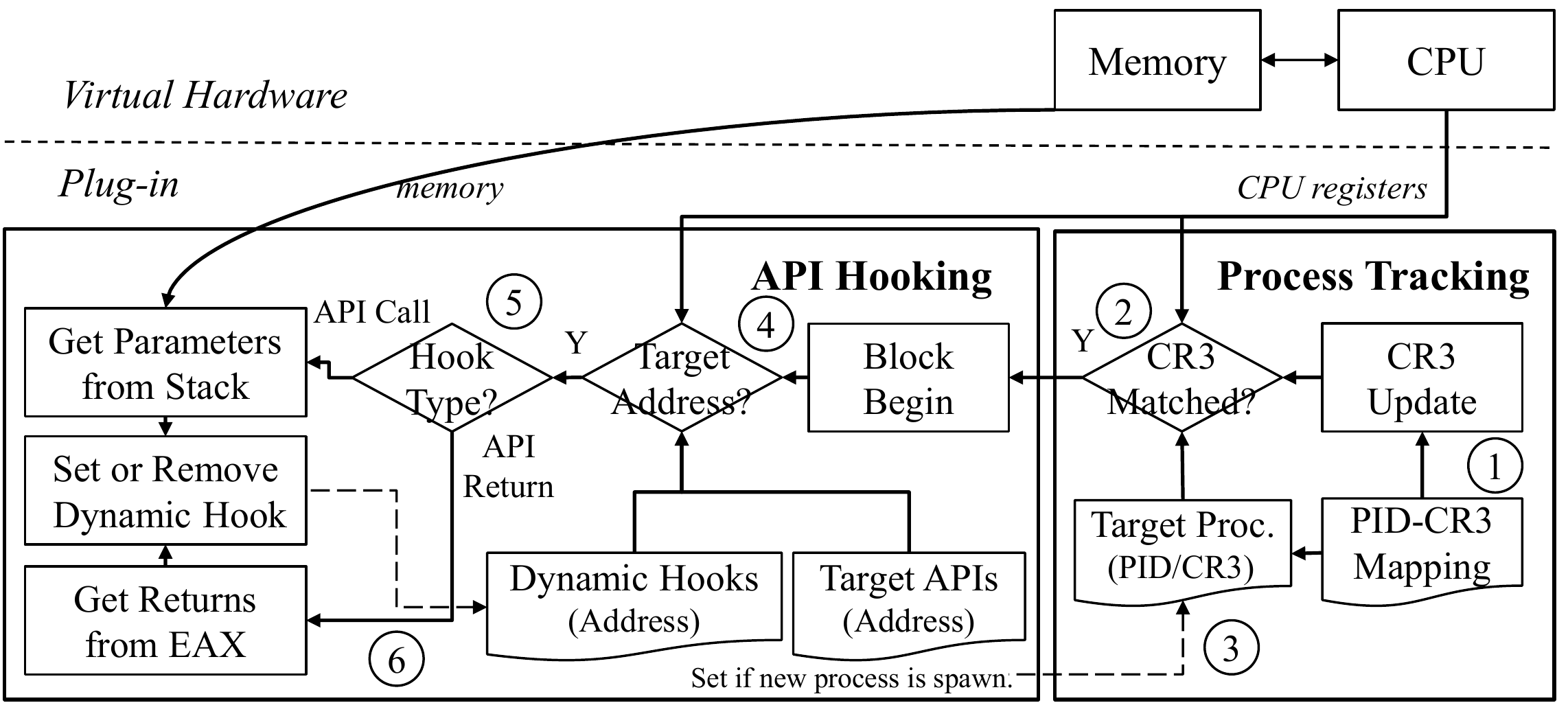}
\caption{The execution flow of the profiling plug-in}
\label{fig:imp}
\end{figure}

\subsubsection{Target Process} The plug-in should be able to identify the target process and track all its spawned child processes. In the guest kernel, each process has a unique identifier, \texttt{PID} to represent the process. When a process is running, the process has a unique address base stored in the CPU register, \texttt{CR3} (which is used for translating virtual address to physical address). The profiling system acquires the \texttt{CR3} value and creates a mapping (step 1 in Fig. \ref{fig:imp}) between the \texttt{CR3} value and the \texttt{PID}. In this case, when the target process is about to running on the virtual CPU, we can start the profiling process.

Since QEMU runs as an emulator, it can check out every instruction before execution. In order to reduce overhead, in VMP, a check-out function is designed that it is executed only before the virtual CPU switches to the next process (i.e., changes the \texttt{CR3} value). If the next process is one of our target processes, the Process Tracking module is activated (step 2). If the target process spawns a child process by calling a Windows API (e.g., \texttt{CreateProcess} or \texttt{WinExec}), the \texttt{PID} of the newly spawned process will be added to the tracking list (step 3).


\subsubsection{Target API} The plug-in should be able to hook the designated Windows API calls. Before executing the malware in the VM, plug-ins and their call back functions are already loaded and to be invoked when corresponding Windows APIs are called by the malware. Forty-three Windows APIs are hooked in the current VMP system, which is categorized into File, Registry, Process, and Library related APIs, as in Table \ref{tab:api}. Some Windows APIs have variants, e.g., \texttt{CreateFileA} and \texttt{CreateFileW} are variants, that only one of them is listed.

To obtain better performance, when QEMU executes the guest codes, it slices the codes into basic blocks, which is a block of instructions terminated by a jump or virtual CPU state change and QEMU executes the basic block as an unit. When a process calls a Windows API function, the instruction in the basic block is exactly a jump to the first instruction of the API function. Hence, the VMP checks if the address of the next instruction (stored in \texttt{EIP} register) is any of the address of our hooked APIs (step 4) before executing a basic block. If the address matches (which means the next instruction will jump to one of the hooked API), then the corresponding call back function is invoked.

\begin{table}
\caption{List of Hooked Windows APIs}
\label{tab:api}
\centering
\begin{tabular}{llll}
\hline
File & Registry & Process & Library \\
\hline
CreateFile & RegCloseKey & CreateProcess & LoadLibrary\\
ReadFile & RegQueryValue & CreateProcessInternal &\\
WriteFile & RegOpenKey & OpenProcess &\\
DeleteFile & RegCreateKey & ExitProcess & \\
CopyFile & RegDeleteKey & WinExec\\
CloseHandle & RegSetValue & CreateRemoteThread &\\
& RegEnumValue & & \\
\hline
\end{tabular}
\end{table}

\subsubsection{Runtime Value Retrieval}
When the CPU jumps to the first instruction of the hooked API function, the call stack already stores the input parameters of this call pointed by the \texttt{ESP} (Extended Stack Pointer) register (step 5). The profiler then gets the parameters from the stack. The number of parameters and data types must be pre-specified in the corresponding call back functions to correctly decode the memory stack. In addition, some Windows APIs use a \textit{handle} to represent a resource, e.g., a file handle (\texttt{HANDLE} \texttt{hFile}) or a registry handle (\texttt{HKEY} \texttt{hKey}). They are basically integers that are meaningless for postmortem analysis. Hence, the VMP system also maintains a mapping table to map resource handles to their original resources, i.e., a file name or a registry key name, when they are opened at the first time.

The VMP profiling system, at the end of call back function, also reads the return address of this API function from the stack, and stores it in an internal monitoring list, called Dynamic Hooks. We will also hook this address for retrieving the return values of this API call. Because after the codes of this API are executed, CPU will jump back to this return address for executing next instruction. At that moment, the return values of the hooked API is at the \texttt{EAX} register and the call-by-reference parameters is in the stack (step 6). That is how do we obtain the return values. Then, we remove this return address from the Dynamic Hooks and records all these parameters and return values in the profile. Finally, the CPU continues to execute the next basic block.

In this way, the high-level semantic information (e.g., Windows API names and human-readable parameters) is obtained by bridging certain low-level information (e.g., register and call stack values). Take a Windows API call \texttt{CreateFile} for example (see Fig. \ref{fig:xml}), to bridge the semantic gaps, the profiling plug-in needs to 1) locate the memory address of the \texttt{CreateFile} API binary codes of the target process, 2) monitor the virtual CPU's \texttt{EIP} register to trigger the hook callback function, 3) translate the binary values in the call stack to human readable values (e.g., \texttt{0x00000010} at the fifth byte from call stack base is translated to \texttt{CREATE\_ALWAYS}), 4) perform dynamic hook for each API call to retrieve the return values, and 5) maintain an internal resource-handle mapping table, where the file names and the corresponding file handles are stored. Thus, the plug-in can output the actual file name (or registry key) in the profile, rather than a handle value. Consequently, a human-readable, high-level, and meaningful behavior profile for each process is populated, as shown in Fig. \ref{fig:xml}.

\section{Malware Behavior Analysis}
\label{sec:analysis}

In this paper, a malware profile is composed of an API execution sequence that many analysis methods can be used to analyze the similarity \cite{Mount} between two API execution sequences for malware family grouping. It is noted that a malware family usually contains many variants, which may be due to the reasons of bug fixing, adding new features, modifying and/or shuffling codes to avoid being captured by an anti-malware engine, etc. However, the variants still inherit intrinsic behaviors from the original codes.

\subsection{Profile Similarity}
The profile of Morstar in Fig. \ref{fig:xml} shows that a malware needs to access specific resources, such as files, Windows registries, and libraries, to accomplish its malicious task. Those resource accessed by the API calls may form the intrinsic behaviors. Hence, the similarity function considers each distinct API with its accessed resources and parameters as an element for similarity calculation. To quantify the similarity of malware, the Jaccard distance, $d_{J}(X, Y)$, defined as in Eq. \ref{eq:j}, is used in this paper to calculate the similarity of a pair of profiles $X$ and $Y$. 

\begin{equation}
\label{eq:j}
  d_{J}(X, Y) = 1 - \frac{|X \cap Y|}{|X \cup Y|}
\end{equation}

$|X \cap Y|$ means the number of elements common in both $X$ and $Y$, and $|X \cup Y|$ means the number of distinct elements in the union of $X$ and $Y$. Let $D$ represent the Jaccard distance matrix that every element $D_{i,j}$ = $d_{J}(i,j)$ is the Jaccard distance of profiles $i$ and $j$. The range of Jaccard distance is between zero (i.e., $X$ and $Y$ are statistically identical) to one (i.e., $X$ and $Y$ do not have any Windows API in common).

\subsection{Malware Behavior Clustering}

To demonstrate the family relationship between malwares, we take the Jaccard distance matrix ($D$) to perform the Unweighted Pair Group Method with Arithmetic Mean (UPGMA) \cite{upgma}, listed in Algo. \ref{algo:UPGMA}, which is a bottom-up hierarchical clustering method, to construct the phylogenetic tree. Both Jaccard distance calculation and the UPGMA algorithm are very efficient, whose computational complexity is $\mathcal{O}( n ^ 2)$ for a naive implementation.

\begin{algorithm}[ht]
 \KwData{A symmetric distance matrix $d$ of size $n \times n$}
 \KwResult{A cluster $C$}

 \ForEach{i in n}{
 Assign $i$ to its own cluster $C[i]$\;
 Create a one-leaf node for $i$ at height 0\;
 }

 \While{the number of current cluster $>$ 1}{
  Find 2 clusters $i$ and $j$ with minimal distance $d[i][j]$\;
  Define a new cluster $k$ by $C[k] = C[i] \cup C[j]$\;
  Define a new node $k$ with children nodes $i$ and $j$\;
  Place new node $k$ at height $d[i][j]$\;
  Add $k$ to current clusters and remove cluster $i$ and $j$\;
  Re-calculate the new distance matrix $d$;
  \tcc{d[x][k]=(d[x][i]+d[x][j])/2}
 }

 \caption{The UPGMA Algorithm}
 \label{algo:UPGMA}
\end{algorithm}

\begin{figure}[tbp]
\centering
\includegraphics[width=0.48\textwidth]{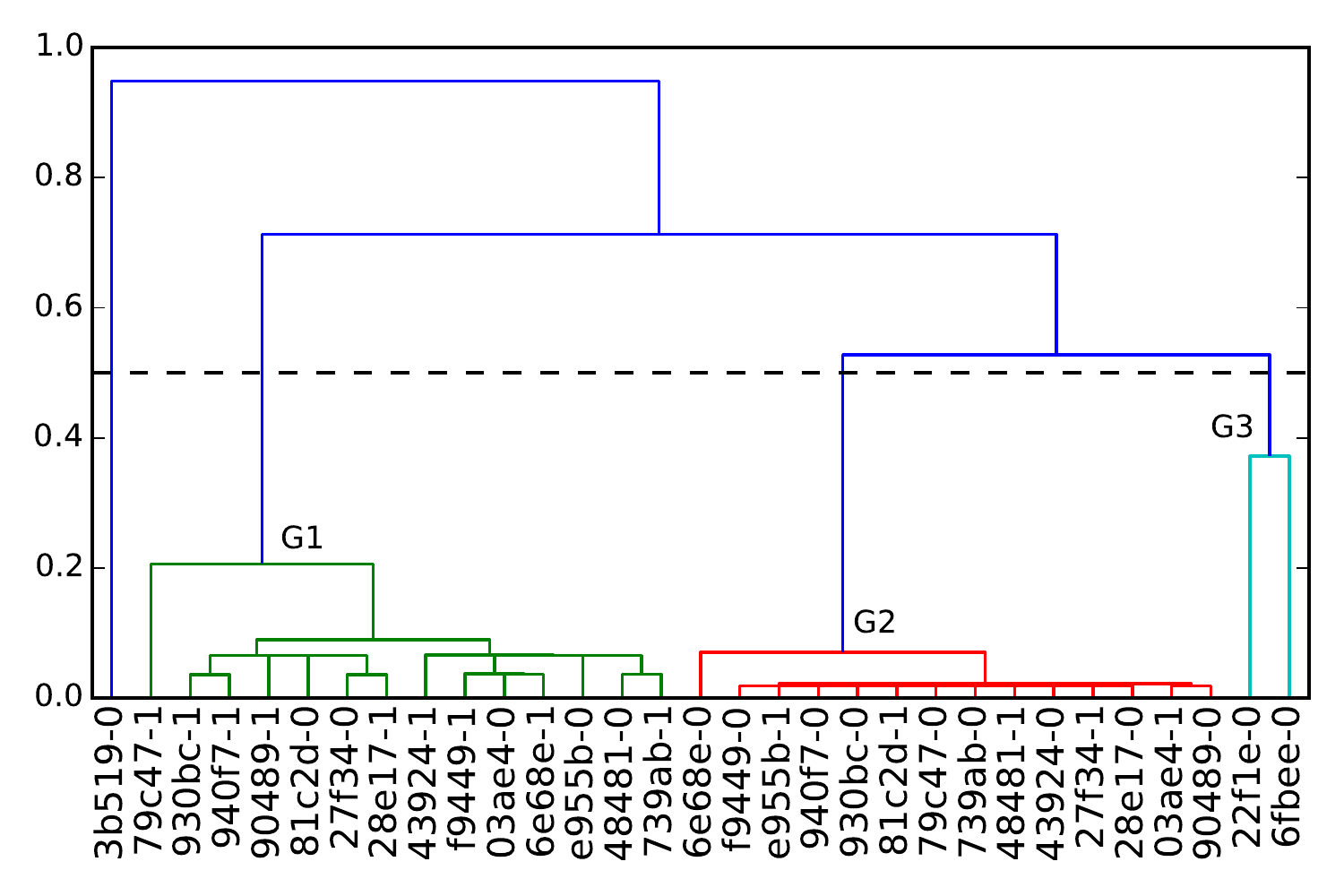}
\caption{The phylogenetic tree of malware family Morstar.}
\label{fig:phy_mor}
\end{figure}

The output of the UPGMA algorithm is a phylogenetic tree. Fig. \ref{fig:phy_mor} shows the phylogenetic tree of Morstar family with 17 malware variants (identified by NCHC \cite{nchc} report) and 31 spawned processes. In VMP, each malware is given a name composed of a hex number and a numerical index. The y-axis of the phylogenetic tree is the Jaccard distance between each pair of malwares or malware behavior groups. For instance, the Jaccard distance of \texttt{22f1e-0} and \texttt{6fbee-0} of $G_{3}$ is 0.37. The shorter distance means higher similarity. In Fig \ref{fig:phy_mor}, if the distance threshold is set to 0.5, the tree can identify three major groups $G_{1}$, $G_{2}$ and $G_{3}$ in this family. More experiment results and discussions of threshold settings are presented in Section \ref{sec:eval}.



\subsection{Malware Behavior Characteristics}





Each group of the phylogenetic tree has unique behavior, which can be ascertained by a collection of API calls. For a group $G$, let $C(G)$ be the common set of APIs with parameters and return values shared by all its children. In practice, an endurance value $\alpha<1$ is set, such that when more than $1 - \alpha$ of its children have the API call (with parameters and return values), the API call is included in $C(G)$. Let $P$ be $G$'s parent in the phylogenetic tree. Then we define the distinct characteristics of $G$, $D(G)$, as all the APIs in $C(G)$ but not in $C(P)$, denoted as follows.

\begin{equation}
\label{eq:g}
  D(G) = C(G) \setminus C(P)
\end{equation}

We then generate the characteristics of each group of the phylogenetic tree. When a new malware is captured, these characteristics can be used to classify the new malware into one of the behavior group with the highest similarity against its $D(.)$ in the phylogenetic tree or none of the groups if the captured malware is not similar to any group. A qualitative explanation of the behavior difference from the security perspective will be given in Section \ref{sec:eval}.

\begin{table*}
\caption{The statistics of API calls of malware families and benign programs}
\label{tab:4cats}
\centering
\begin{tabular}{lrrrcrrrrcrrrrcr}
\hline
 & \multicolumn{3}{c}{File} & & \multicolumn{4}{c}{Registry} & & \multicolumn{4}{c}{Process} & & \multicolumn{1}{c}{Library}\\
\cline{2-4} \cline{6-9} \cline{11-14} \cline{16-16}
 & Create & Copy & Delete & & Query & Create & Delete & SetValue & & Create & Open & Thread & WinExec & &  LoadLibrary\\
\hline
Korgo & 8.6 & 1.0 & 0.0 & & 131.3 & 12.0 & 1.0 & 10.7 & & 1.4 & 2.5 & 1.0 & 1.0 & & 24.0\\
Pinfi & 16.5 & 1.2 & 0.1 & & 189.6 & 19.5 & 1.2 & 20.2 & & 2.4 & 1.9 & 1.0 & 1.2 & & 25.9 \\
Sality & 12.3& 0.8 & 2.3 & & 466.3 & 18.1 & 125.8 & 388.0 & & 0.4 & 9.8 & 2.4 & 1.0 & & 23.0\\
Virut & 7.0 & 1.0 & 0.0 & & 119.1 & 16.7 & 1.0 & 14.4 & & 1.0 & 2.2 & 1.0 & 1.0 & & 23.8\\
\hline
IE & 197.0 & 0.0 & 6.0 & & 1015.0 & 34.0 & 0.3 & 40.0 & & 0.0 & 3.0 & 0.0 & 0.0 & & 42.0\\
Chrome & 93.0 & 1.0 & 10.0 & & 409.0 & 94.0 & 0.0 & 8.0 & & 1.0 & 3.0 & 0.0 & 0.0 & & 46.0\\
MSN & 14.0 & 0.0 & 0.0 & & 350.0 & 19.0 & 0.0 &19.0 & & 0.0 & 0.0 & 0.0 & 0.0 & & 19.0\\
MS Paint & 4.0 & 0.0 & 0.0 & & 188.0 & 21.0 & 0.0 & 0.0 & & 0.0 & 0.0 & 0.0 & 0.0 & & 11.0\\
\hline
\end{tabular}
\end{table*}

\section{Evaluations}
\label{sec:eval}

\subsection{System Platform}

The platform uses off-the-shelf machines (Intel i7-3770S 3.1 GHz CPU with 8 GB RAM, Gigabit Ethernet, and 500 GB hard disk) with 64-bit Ubuntu 12.04 LTS (Linux kernel 3.11.0.26) as host OS. The TEMU/QEMU \cite{temu,temu2} is employed as VMM, and the guest OS is a vanilla Windows XP SP3. The configuration of a VM is a single core CPU at 3.1 GHz with 1GB RAM and 20 GB hard disk.



\subsection{Malware Data Sets}

In the evaluations, two malware data sets are used that both are provided by the National Center for High-Performance Computing (NCHC), Taiwan \cite{nchc}. The first set, 40Bot, was collected in late 2011, consisting of four well-labeled botnet families (Virut, Sality, Korgo and Pinfi) with 10 variants of each family. The second set, 419Mal, has 272 malwares (which fork 419 processes in total) collected from April 2014 to October 2014 from a collaborated honeynet run by NCHC. From the records of VirusTotal, the first-seen date of the 419Mal malwares were from August 2009 to October 2014. In the evaluation, each malware runs on a VM for 300 seconds to generate its profile (as in Fig. \ref{fig:xml}) that their profile size is up to 334 KB with 57 KB in average.


\subsection{API Calling Statistics and Usages}

40Bot and 4 benign programs (Internet Explorer, Google Chrome, MSN Messenger and MS Paint) are profiled and examined to demonstrate the behavior similarity of each malware family and their differences with benign programs. It is expected that each malware family has its unique characteristics, and has certain distinctions with other families and benign programs. Table \ref{tab:4cats} shows the statistics of average numbers of APIs called by malware families and benign programs. The findings of interesting API usages are elaborated below.

\textbf{Process-related APIs.} \texttt The APIs \texttt{CreateProcess}, \texttt{WinExec}, \texttt{OpenProcess} and \texttt{CreateRemoteThread} fall into this category. It is common for malware to spawn new processes for malicious jobs, using \texttt{CreateProcess} and \texttt{WinExec}. On the contrary, IE only opens new process named \texttt{IEXPLORER.EXE} by using \texttt{OpenProcess}. Their difference can be observed from the process column of Table \ref{tab:4cats}. Malware tends to use \texttt{WinExec} to execute another program, but none of the selected benign programs do so. For example, Korgo uses \texttt{WinExec} with command line parameter: \texttt{C:{\textbackslash}}\texttt{WINDOWS{\textbackslash}}\texttt{system32{\textbackslash}zaegr.exe} to start a malware process. The usage of process-related APIs is quite different in the implementations of malware and benign programs, due to different purposed objective and style of coding practice.

\textbf{CopyFile and DeleteFile.} Most of the malware use \texttt{CopyFile} to make a copy of the malware binary to Windows system folder or temporary folder for later execution. The filenames are usually random names, such as \texttt{vwjop.exe}. Sality deletes several files, which are all temporary files created by itself. For IE and Chrome, they only delete HTTP cookies and temporary HTML files in the browser's temporary folder. While the same API is used by both malware and benign program, the parameters are totally different.
 
\textbf{CreateFile.} The \texttt{CreateFile} function has a parameter \texttt{creationDisposition} with values \texttt{CREATE\_NEW}, \texttt{CREATE\_ALWAYS} or \texttt{OPEN\_EXISTING}. From the observation, malwares usually use the first two values to create files in the system folder for later use, and use the last one to read or execute existing files, while IE only uses \texttt{OPEN\_EXISTING} for reading cookies, cached HTML and font files from its temporary folder and font folder only. Again, the differences are embedded in the parameter values.

\textbf{RegCreateKey, RegSetValue and RegQueryValue.} Malware may modify or add a registry to change the behavior of the infected host, such as adding new service, changing the host name or domain name (as Korgo), changing firewall settings (as Sality), disabling Limited User Account (as Sality), etc. In addition, some malware tries to register itself as a service, so that it will be executed automatically after system reboots. The above operations can be distinguished from the parameters used in the APIs.

\textbf{RegDeleteKey.} Malware tends to delete registry keys to remove system features. For instance, Sality deletes all registry keys under a subkey \texttt{SafeBoot}. Deleting this key prevents the user from booting into Safe Mode, which makes the user unable to fix the infected host.

\textbf{LoadLibrary.} Malware may load third-party library files from a non-system directory, but not the four benign programs.

The above findings and the statistics in Table \ref{tab:4cats} show that malware families have different API calling patterns and the usages of parameters with benign programs, as well as among different families, which motivates further study of the usage of API parameters in malware profiling.


\begin{table*}
\caption{The Jaccard distance and statistics of malware families with Internet Explorer}
\label{tab:jaccard}
\centering
\begin{tabular}{lcccc}
\hline
 & Korgo & Pinif & Sality & Virut \\
\hline
Avg. ${d_{J}(.)}$ & 0.1471 (0.1660) & 0.5080 (0.2055) & 0.5151 (0.2575) & 0.4403 (0.2346) \\
Avg. ${d_{J}(., IE)}$ & 0.8730 (0.0041) & 0.8828 (0.0265) & 0.9289 (0.0388) & 0.8601 (0.0169)\\
Gap & 0.7258 & 0.3748 & 0.4138 & 0.4198 \\
T-Test & 2.09347E-30 & 8.18355E-16 & 8.30675E-14 & 1.96907E-15 \\
\hline
Avg. ${d_{J}(.)}, no\_par$ & 0.0000 (0.0000) & 0.0311 (0.0333) & 0.0535 (0.0593) & 0.0311 (0.0333) \\
Avg. ${d_{J}(., IE)}, no\_par$ & 0.3571 (0.0000) & 0.3700 (0.0197) & 0.3054 (0.0389) & 0.3700 (0.0197)\\
Avg. ${d_{J}(., GC)}, no\_par$ & 0.2857 (0.0000) & 0.3000 (0.0218) & 0.3690 (0.0306) & 0.3000 (0.0218)\\
\hline
\end{tabular}
\end{table*}

\subsection{Family Similarity -- Jaccard Distance}

Table \ref{tab:jaccard} shows the average Jaccard distance and standard deviation among the variants of each malware family in 40Bot and their average distances with IE, denoted as ${d_{J}(.)}$ and ${d_{J}(., IE)}$, respectively. It is clear that the average distance within a malware family is much smaller (i.e., more coherent in their behaviors) than the average distance between a malware family and IE. The execution behaviors of the variants in Korgo are most similar to each other (with the average distance 0.1471) among all the families. On the other hand, the average distance between Sality variants is higher (0.5151); however, it is still sufficient to distinguish the family members from IE (0.9289). The degree of code modification of malware family results in different average Jaccard distance. The p-values of T-test of all malware families with IE reject the false null hypotheses that IE and malware variants belong to the same group.
The bottom rows of Table \ref{tab:jaccard} lists the Jaccard distances of profiles without the parameters and return values of API calls (denoted as no\_par), which shows relatively poor distinction between malwares and both Internet Explorer (IE) and Google Chrome (GC), e.g., 0.3571 between Korgo and IE, and 0.2857 between Korgo and GC. As Korgo and IE (GC) share around 63\% (71\%) Windows APIs in their programs, without considering parameters and return values make them indistinguishable. This shows the drawback to use API names only in distinguishing between software behaviors.

\begin{figure}
\centering
\subcaptionbox{Korgo\label{fig:phy_01}}[.24\textwidth]{\includegraphics[width=0.23\textwidth,clip=true,trim=0mm 0mm 0mm 0mm]{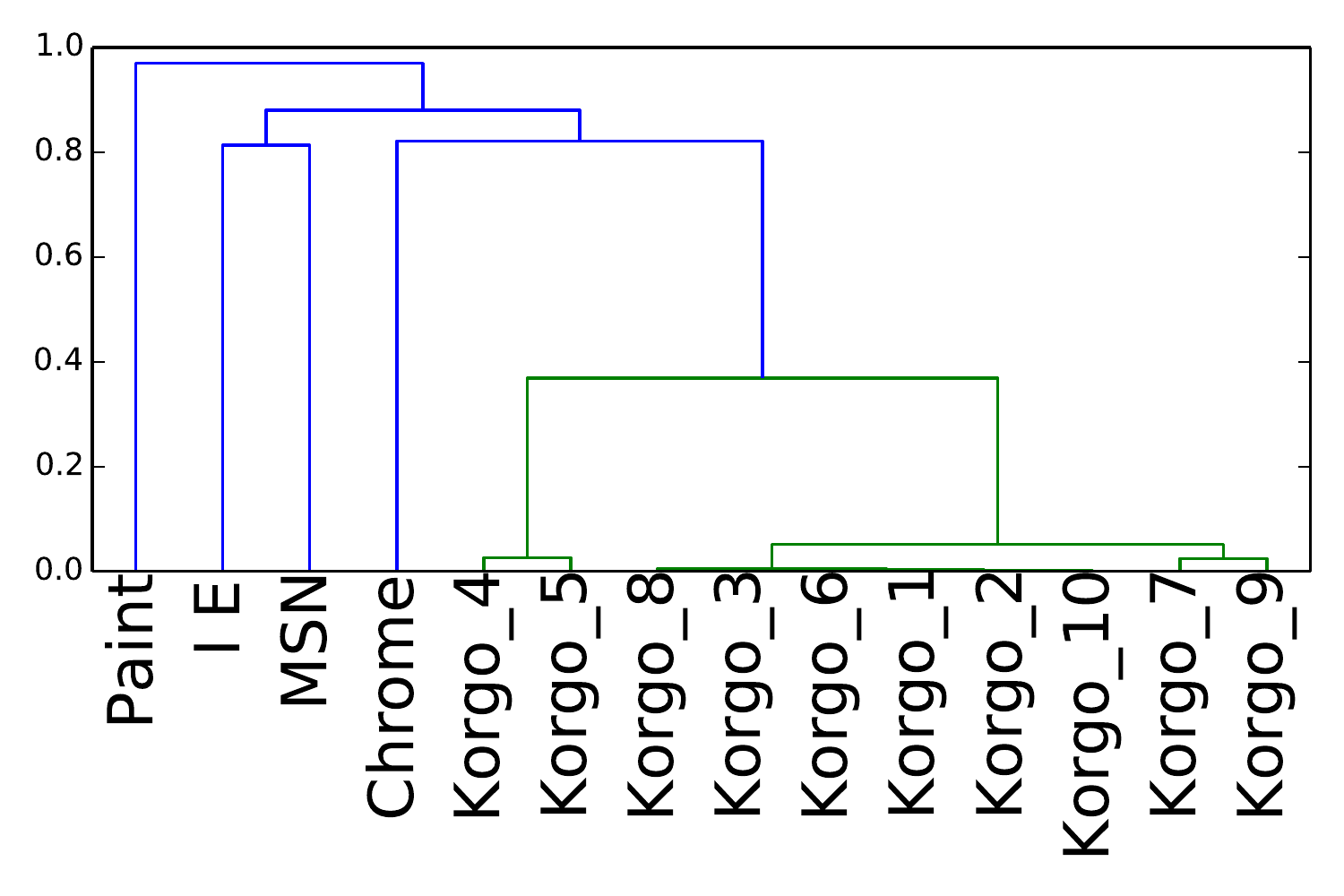}}%
\subcaptionbox{Pinif\label{fig:phy_02}}[.24\textwidth]{\includegraphics[width=0.23\textwidth,clip=true,trim=0mm 0mm 0mm 0mm]{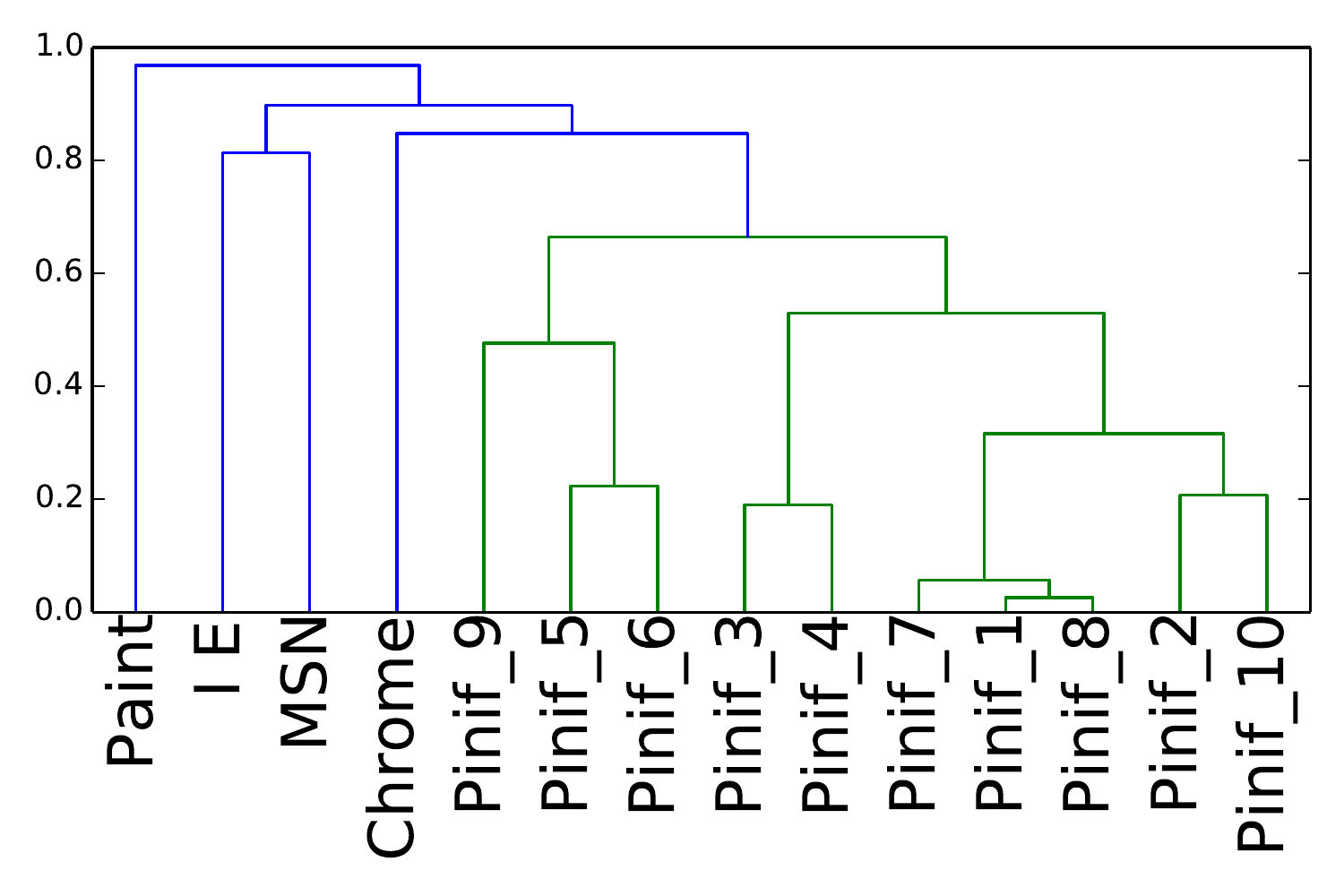}}
\\
\subcaptionbox{Sality\label{fig:phy_03}}[.24\textwidth]{\includegraphics[width=0.23\textwidth,clip=true,trim=0mm 0mm 0mm 0mm]{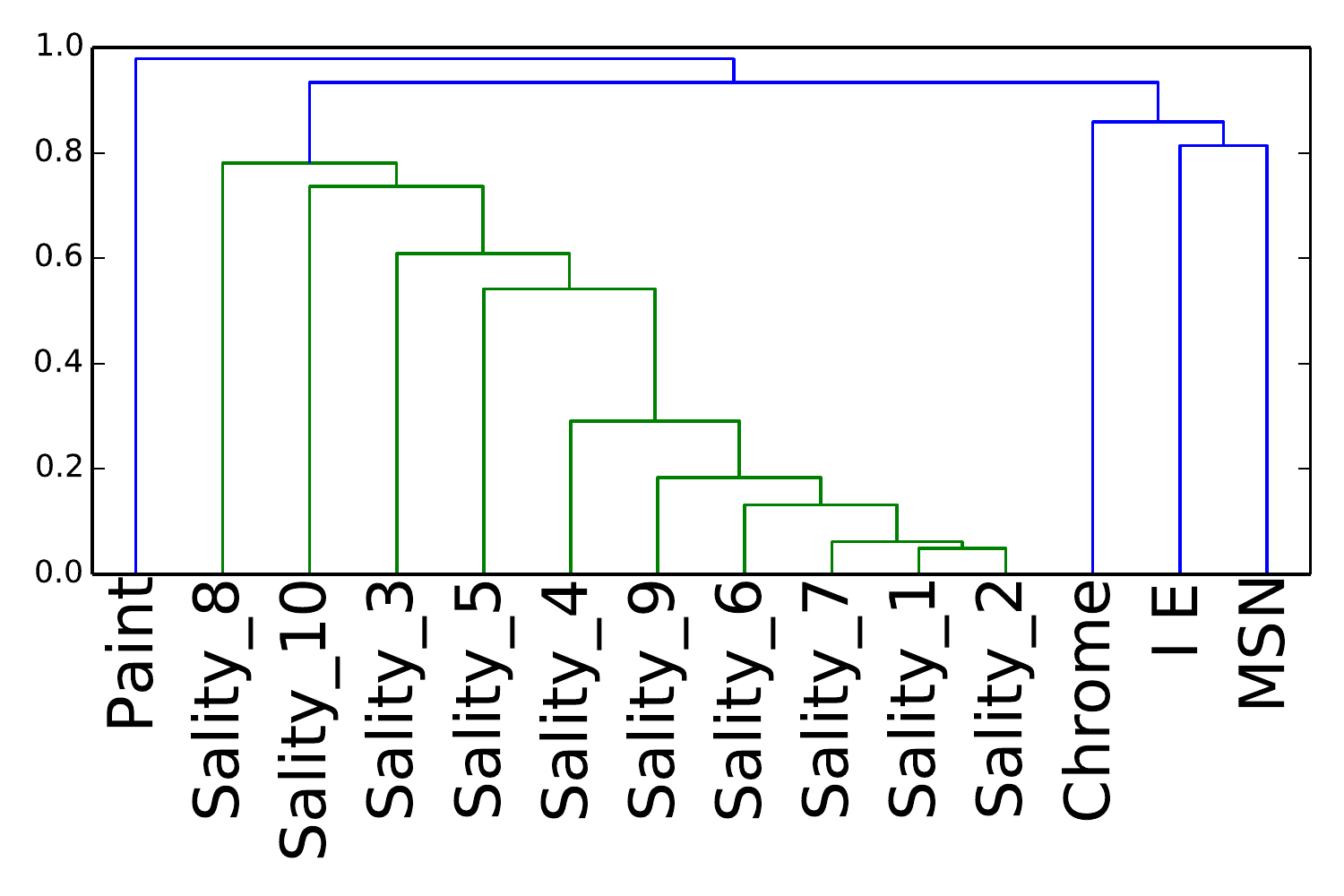}}%
\subcaptionbox{Virut\label{fig:phy_04}}[.24\textwidth]{\includegraphics[width=0.23\textwidth,clip=true,trim=0mm 0mm 0mm 0mm]{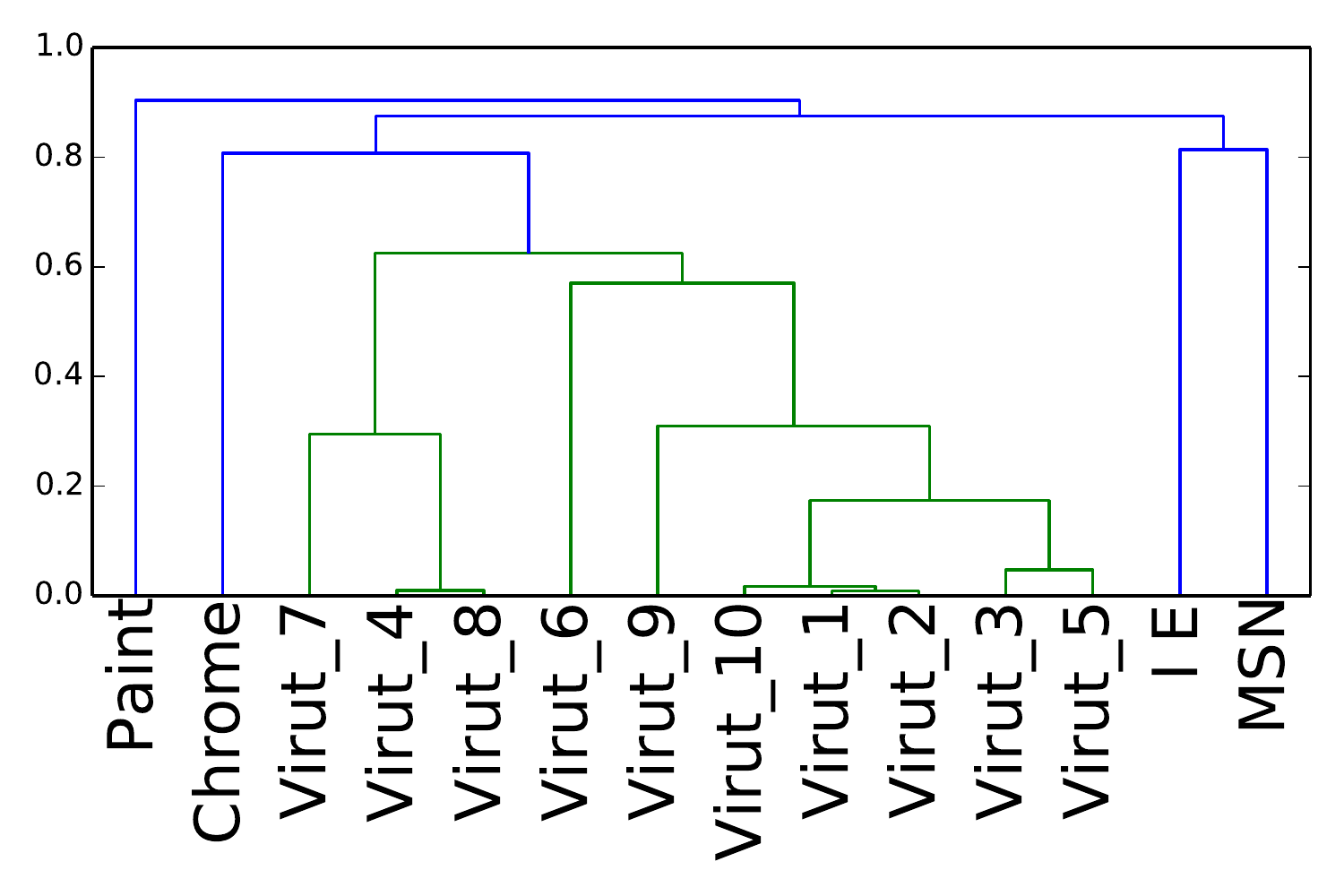}}
\caption{The phylogenetic trees of four malware families with four benign programs.}
\label{fig:phy_40}
\end{figure}

Fig. \ref{fig:phy_40} shows the phylogenetic tree of malware families with the four benign programs to visualize the family construction and their associated Jaccard distances. It is interesting to observe that the benign programs cluster with malwares only with very high Jaccard distances. For instance, in Fig. \ref{fig:phy_40}(a), Chrome clusters Korgo family at Jaccard distance around 0.8. The construction of each malware family tree presents its own evolution pattern (i.e., code alternation history). For instance, Sality family members have a sequence of code alternations, while Korgo family members present two major behavior groups.




%
%

\subsection{Family Similarity -- Phylogenetic Tree}

In this section, an intensive analysis of a malware family Morstar is presented to exemplify the functionality and capability of the proposed scheme. 
Fig. \ref{fig:phy_mor} is the phylogenetic tree of the malware family Morstar with three major groups when a threshold value is set at 0.5. After further investigations, it is observed that $G_{1}$ launches a File Explorer, removes the browsing records, changes some Internet settings, and sets up an auto-execution file on the system drive, $G_{2}$ checks the system services in the host, makes a copy of the malware binary and executes it, and $G_{3}$ is similar to $G_{2}$, but the execution trace is much shorter that indicates unsuccessful attacks during the period of profile generation. 

The phylogenetic tree faithfully shows the intrinsic behaviors and relationships inside each subtree and among subtrees. By referring to published documents, Morstar is an adware/trojan which is bundled with other software installers. Apparently, the variant \texttt{3b519-0}, the leftmost one in Fig. \ref{fig:phy_mor}, is an exception case in this family. The Jacquard distance is large enough (0.9488) to determine that it is an outlier. It is not clear why some detection engines classify this variant into Morstar family that it impels to study the discrepancy among detection engines.


\begin{table}
\caption{The significant characteristics of Morstar variants\label{tab:intersection}}
\begin{center}
\begin{tabular}{lcccp{4.1cm}}
\hline
Group & Size & $|C(G)|$ & $|D(G)|$ \hspace{1mm}& Samples of Distinct Characteristics\\
\hline
\noalign{\smallskip}
$G_{1}$ & 14 & 610 &490 & \parbox[t]{4cm}{RegCreateKey: P3Sites, P3Global;\\RegSetValue: Shell Folders;\\RegSetValue: ProxyEnable;\\RegCreateKey: CmdMappingIt} \\
$G_{2}$ & 14 & 304 & 184 & \parbox[t]{4cm}{CreateProcessInternal;\\RegCreateKey: Policies;\\LoadLibrary: netapi32;\\RegCreateKey: Blocked} \\
$G_{3}$ & 2 & 120 & 0 & \parbox[t]{4cm}{$\emptyset$} \\
\hline
\end{tabular}
\end{center}
\end{table}


Table \ref{tab:intersection} shows the statistics and characteristics of the three major groups in Fig. \ref{fig:phy_mor}. For example, $G_{1}$ has 14 malware processes, and 610 Windows API calls are common among the members in $G_1$. Within these 610 API calls, there are 490 calls are identified as the distinct characteristics of $G_{1}$ by Eq. \ref{eq:g}. 
The distinct characteristics of the malware groups in Table \ref{tab:intersection} match the observations. Take $G_{1}$ as an example that it creates two malicious registry keys, \texttt{P3Sites} and \texttt{P3Global} to modify the setting of IE. It also changes the location of folder redirection, e.g., Desktop, Start Menu, and My Document, by modifying the registry of \texttt{Shell Folders}. It disables the HTTP proxy by modifying registry key \texttt{ProxyEnable} to avoid being monitored. It modifies the registry \texttt{CmdMappingIt} to change the buttons on the Command Bar of IE. The above APIs are discovered as the distinct characteristics in VMP.

For $G_{2}$, one of distinct characteristics is that it invokes \texttt{CreateProcessInternal} API to create a process from Windows \texttt{TEMP} directory. It later changes the settings of Windows Explorer by modifying registry \texttt{Policies}. The variants in this group load a specific library, \texttt{netapi32}, which is not loaded by the variants in $G_{1}$. It also blocks many shell extensions in Windows system that changes the behavior of Windows systems.

$G_{3}$ has only two variants that have 120 common API calls. From the manual analysis, besides querying registries, the variants only create one file and one Windows registry, which are considered as unsuccessful attacks during the period of profile generation.

\subsection{Pairwise Classification Score}

Each anti-malware detection engine has their own schemes to name a malware and to define a malware family. Unfortunately, the malware families and their construction rationales have great discrepancies among different engines, which demands a metric to evaluate the behavior classification capability and accuracy of different anti-malware detection engines. Some websites, such as VirusTotal \cite{VirusTotal}, analyze and consolidate malwares detected by detection engines to provide a list of malwares and their detection names returned by each engine. As we can see in VirusTotal, a single malware (identified by its hash value) may have different given names by different engines.


\begin{figure*}[t]
\centering
\includegraphics[width=0.99\textwidth,clip=true,trim=37mm 134mm 26mm 109mm]{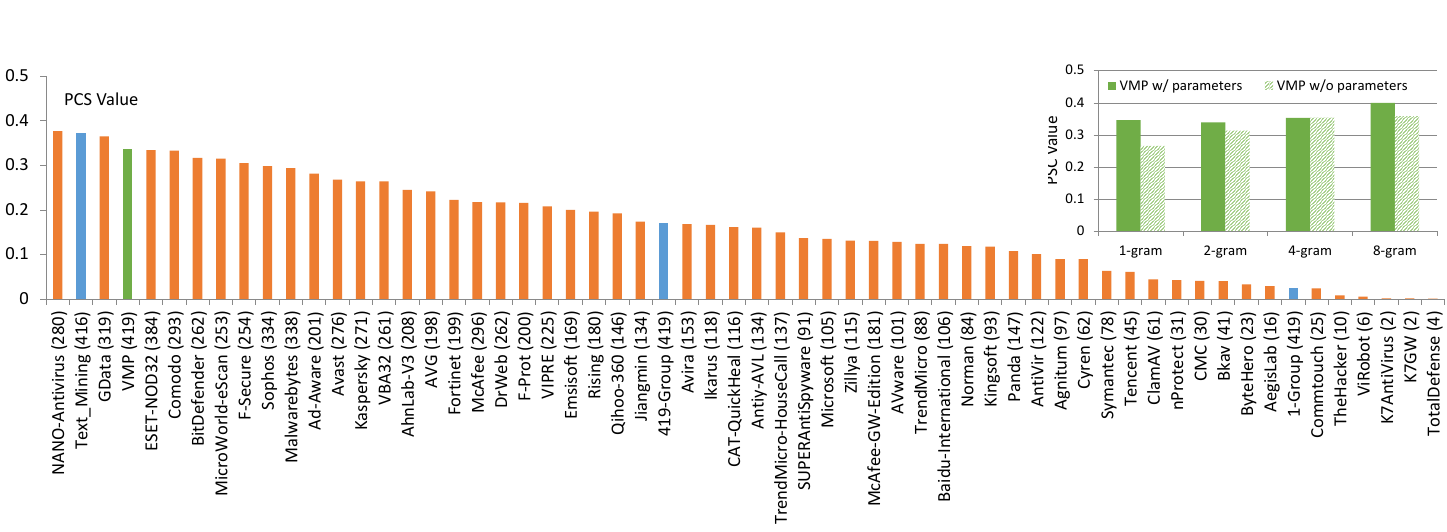}
\caption{The Pairwise Classification Scores of detection engines.}
\label{fig:pcs}
\end{figure*}




Since there has no ground truth of how to define a malware family, we leverage the peer voting concept to design a voting mechanism to determinate which detection engine has 'better' family grouping capability. We propose a metric, Pairwise Classification Score (PCS), for the evaluation of goodness of detection engines against the proposed VMI-based behavior profiles while using them in family grouping. The intuition of PCS adopts the idea of crowd intelligence that a good detection engine should be acclaimed by other engines. Assume there are $n$ malwares, denoted as $M_{i}$, and $m$ detection engines, denoted as $E_{x}$. In this case, we can construct a table with $n \times m$ cells, and for a cell $(i, x)$, it contains the detection string of $M_{i}$ by $E_{x}$.

There is an important assumption of PSC that we cannot directly use cell $(i, x)$ and $(i, y)$ for peer voting, since $E_{x}$ and $E_{y}$ may use different malware naming scheme. Take the malware sample in Fig. \ref{fig:xml} for example, Avira names it as `APPL/Firseria.A.15', Kaspersky names it as `Win32.Morstar.ba' and Sophos calls it `Solimba Installer'. In this case, while voting, we take a pair of $M_{i}$ and $M_{j}$ as the voting basis. To evaluate the goodness of $E_{x}$ by using $E_{y}$, we consult the cells at $(i, x)$, $(j, x)$, $(i, y)$ and $(j, y)$. If $E_{x}$ says $M_{i}$ and $M_{j}$ belong to the same family (i.e., cell $(i, x)$ and $(j, x)$ share the same family name under $E_{x}$'s naming scheme), and $E_{y}$ also says so (i.e., cell $(i, y)$ and $(j, y)$ share the same family name under $E_{y}$'s naming scheme), then $E_{y}$ gives $E_{x}$ a positive point; otherwise negative. On the other hand, if $E_{x}$ says $M_{i}$ and $M_{j}$ are not in the same family, and $E_{y}$ also says so, then a positive point is given to $E_{x}$ by $E_{y}$ as well.

Let $O_{x}(i)$ be the malware family of $M_{i}$ given by $E_{x}$ and the value is NULL if $E_{x}$ fails to detect this malware. Let $I_{x}(i,j)$ be an indicator function, which indicates if two malwares, $M_{i}$ and $M_{j}$, are considered to be in the same family by $E_{x}$, which is defined as below.

\begin{equation}
\label{eq:in}
  I_{x}(i,j) = \left\{
  \begin{array}{l l}
   +1&\text{if $O_{x}(i)=O_{x}(j)$}\\
   0&\text{if $O_{x}(i)$ or $O_{x}(j)$ is NULL}\\
   -1&\text{if $O_{x}(i) \neq O_{x}(j)$}
  \end{array} \right.
\end{equation}

Then, the Pairwise Classification Score of a detection engine $E_{x}$, denoted as $PCS_x$, is determined by referencing all $m$ engines on all $C(n,2)$ pairs of malwares. Let $W_{x}$ be the weight for $E_{x}$ (and in our work, it is calculated by the portion of malwares detected by $E_{x}$). $P_{x}(y)$ is the approval rate of engine $x$ from engine $y$. Intuitively, the higher $PCS_{x}$ indicates a more recognized engine. Note that since naming schemes are different engine by engine, the proposed PCS method is a much more proper way to evaluate malwares classification results among multiple engines.

\begin{equation}
\label{eq:pcs}
PCS_{x} =  \frac{1}{m} \times W_{x} \times \sum_{y=1}^{m} P_{x}(y)
\end{equation}

where

\begin{equation}
\label{eq:w}
\begin{split}
W_{x} & = 1 - P\left( O_{x}(i) = NULL \right) \\
\end{split}
\end{equation}

\begin{equation}
\label{eq:pxy}
\begin{split}
P_{x}(y) & = P\left( I_{y}(i,j) = +1 \mid I_{x}(i,j) = +1 \right)\\
& + P\left( I_{y}(i,j) = -1 \mid I_{x}(i,j) = -1 \right)\\
\end{split}
\end{equation}


\subsection{Detection Engine Performance Comparisons}

Fig. \ref{fig:pcs} shows the PCS values of the proposed VMP system, 56 engines listed by VirusTotal plus three types of detection engines (N-Group, N-Gram, and Text\_Mining) built for comparisons. The value in the parenthesis of each engine in Fig. \ref{fig:pcs} is the number of malwares detected by the engine. Note that the PCS values do not imply high detection numbers and vice versa. The N-Group engine clusters the malwares into $N$ groups. The N-Gram engines consider $N$ consecutive APIs as a behavior unit in the profile (that will affect the Jaccard distance calculation). The Text\_Mining engine generates a classifier using the text description of each malware.

\textbf{N-Group.}
In the comparison, two engines (1-Group and 419-Group) are built as two extremes without any knowledge of malware. 1-Group clusters the 419 malware processes into one big family and achieves PCS = 0.0261, and 419-Group generates a family for each malware process and achieves PCS = 0.1703. 

\begin{figure}
\centering

\subcaptionbox{N-Gram\_N
\label{fig:p0ie}}[.24\textwidth]{\includegraphics[width=0.24\textwidth,clip=true,trim=0mm 0mm 0mm 0mm]{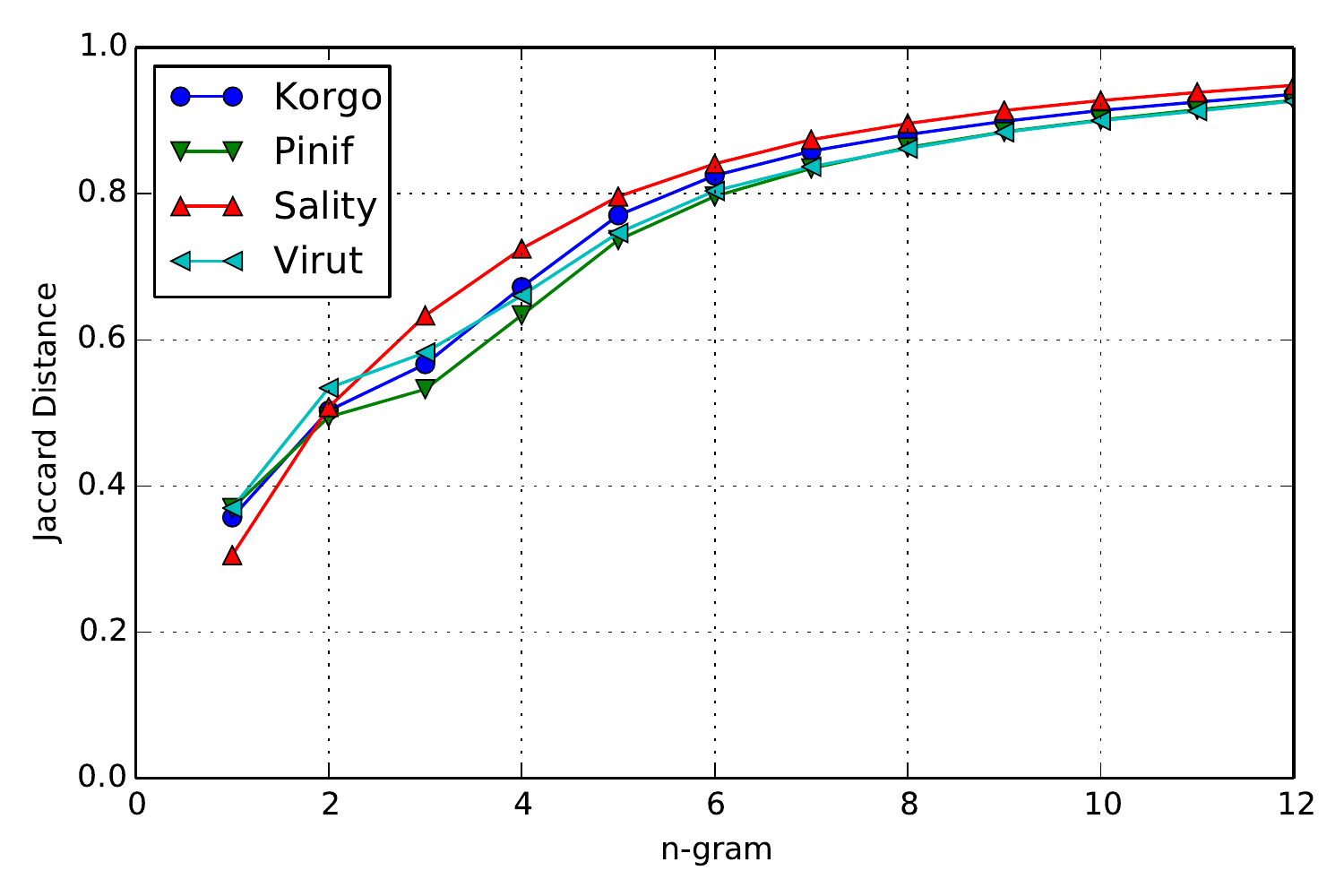}}
\subcaptionbox{N-Gram\_Y
\label{fig:p1ie}}[.24\textwidth]{\includegraphics[width=0.24\textwidth,clip=true,trim=0mm 0mm 0mm 0mm]{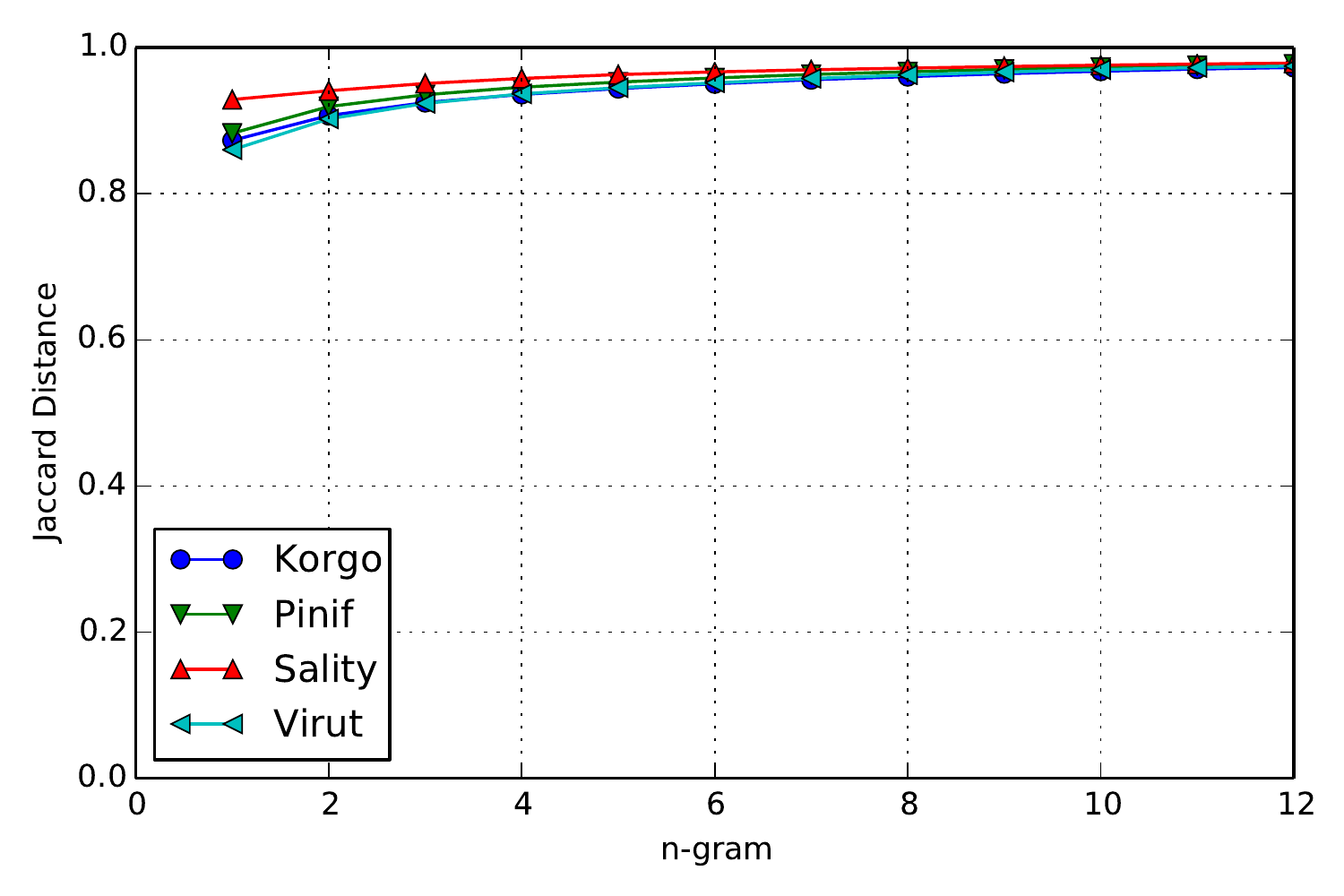}}%
\caption{Jaccard Distances of Malwares with benign software using N-Gram}
\label{fig:p01ie}
\end{figure}

\textbf{N-Gram.} Inspired by Forrest's work \cite{sense}, we study the effectiveness of using $N$ consecutive API calls (denoted as N-Gram) in malware family grouping that the case considers parameters and return values is denoted as N-Gram\_Y, and the case does not consider parameters and return values is denoted as N-Gram\_N. (Note the proposed VMP system is 1-Gram\_Y.) Fig. \ref{fig:p01ie} shows the Jaccard distance between benign software and four malware families (of 40Bot) from 1-Gram to 12-Gram. Note that the obvious problem of N-Gram approach is high computation overhead that N-Gram costs $\mathcal{O}(N)$ times of 1-Gram and the space complexity is $\mathcal{O}(d^N)$, where $d$ is the number of hooked APIs in VMP.

In Fig. \ref{fig:p0ie}, it shows that the Jaccard distances of N-Gram\_N with benign software going up with the increase of $N$, and get saturated when $N$ is larger than 10. Fig. \ref{fig:p1ie} demonstrates the benefit of considering parameters and return values (i.e., N-Gram\_Y) that 1-Gram\_Y already achieves the Jaccard value as high as 9-Gram\_N.

In Fig. \ref{fig:pcs}, it also presents the results of N-Gram\_Y and N-Gram\_N with threshold = 0.3. Their PCS values are 0.3462, 0.3513, 0.3534, and 0.3996 for N-Gram\_Y, when $N$= 1, 2, 4, and 8, respectively, and 0.2658, 0.3127, 0.3534, and 0.3580 for N-Gram\_N. Even for 1-Gram\_Y, the PCS value is on par with 8-Gram\_N, which shows the benefit of considering parameters and return values.

\textbf{Text\_Mining.} As every detection engine give its description of each detected malware manually, the Text\_Mining engine collects the descriptions from all engines as its description of the malware, and applies the text mining algorithm modified from ACIRD \cite{acird} to generate malware families. A list of stop word is prepared, such as ``Win32'', ``Variant'', ``TROJ\_GEN'', special symbols and strings, and punctuations, to remove them from the malware description. 

Once the bag-of-word frequency matrix is generated, cosine similarity is calculated between each pair of malwares, $M_i$ and $M_j$. If their cosine similarity is smaller than a pre-specified threshold value, we consider $M_i$ and $M_j$ are in the same group, i.e., $I_{Text\_Mining}(i,j) = +1$ in Eq. \ref{eq:in}. With the threshold value 0.7, the Text\_Mining engine achieves its best PCS at 0.3736. It is not surprising that the Text\_Mining approach performs well since it takes advantages of the descriptions of all 56 detection engines prepared by the security experts. On the other hand, the Text\_Mining approach highly depends on the quality of manual descriptions.  

\textbf{VMP Performance.} The VMP system (i.e., 1-Gram\_Y) achieves a high PCS = 0.3462, just next to two well-managed detection engines. The major reason is that a malware may spawn several child processes that VMP considers each process independently by its behavior in family, while all other detection engines consider the child processes in the same malware family with their parent. Although we believe VMP works correctly in the aspect of behavior clustering, most other engines may disagree and give negative evaluations.


\subsection{10-fold Testing} 
In this subsection, a 10-fold testing is performed to examine the performance of VMP in classifying unseen malware. For each test, 42 malwares are randomly selected from 419Mal as the testing data (as the unseen malwares), and the rest 377 malwares are used as the training data. The test repeats for 30 times in this experiment. 

First, the training data is used to build the corresponding phylogenetic tree, groups and their distinct characteristics. Then, every testing malware is fed into the phylogenetic tree to calculate the most similar group by comparing the distinct characteristics of a group and the testing malware. If the grouping outcome is the not same with the original VMP phylogenetic tree of 419Mal, the grouping result is negative; otherwise, positive.

Table \ref{tab:10-fold} shows the results of 10-fold testing for different thresholds. For instance, if the threshold is set to 0.3, there are 81 groups are identified in the constructed phylogenetic tree and only 1.9048\% malware processes (0.8 out of 42) are assigned to a wrong group. There are 11.8254\% (4.9666 out of 42) of testing malware processes that form a single member malware group in the original phylogenetic tree and in the testing phase, these malwares cannot be classified into any group as it is supposed to be. As expected, if the Jaccard distance threshold is low, then the number of groups increases and their characteristics get strict and precise. In this case, the negative rate becomes low with many one-member groups and the opposite for high threshold.

\begin{table}
\caption{10-fold Testing on 419Mal}
\label{tab:10-fold}
\centering
\begin{tabular}{crrr}
\hline
Threshold & Groups & Negative Rate (\%) & Single-member Rate (\%) \\
\hline
0.2 & 92 & 0.7143 & 13.4921\\
0.3 & 81 & 1.9048 & 11.8254\\
0.4 & 62 & 3.4127 & 9.7619\\
0.5 & 54 & 7.4603 & 6.6666\\
\hline
\end{tabular}
\end{table}

\section{Related Work}
\label{sec:related}
\subsection{Profiling Subject}

Bayer et al. \cite{view} published an analysis report based on the malicious code samples (total 901,294 unique samples) that were collected by Anubis \cite{dynamic} during 2007--2008. They provide an overview of observed behavior among these samples. The file system, registry, network, GUI, and process-related activities are the most commonly shared behavior by these malwares.

There are many ways in which system call data \cite{Calls} could be used to characterize the behavior of programs, such as enumerating sequences\cite{sense,Sequences}, frequency-based method and data mining approach \cite{mining}. Usually, the result can be used to compare the execution behavior of a known malware against that of a set of benign programs\cite{specifications}. If any pattern observed is out of the predefined model, an anomaly is detected. Some literature \cite{profile,BehaviorSpyware} also focus on Windows API calls.

In our work, we come out the similar result but we focus ons Windows API to populate malware profile. Moreover, we recorded the parameters and the return values with the Windows API calls that these malwares use. Our experiments show that the Windows AIP calls may be the same, but the parameters could reveal different semantics.

Some researchers focus on the higher-level semantic-meaningful behavior of different malware, e.g., botnet network connection pattern or e-mail propagation activity. BotHunter \cite{BotHunter} constructs a bot infection dialog model by a set of loosely ordered communication. BotSniffer \cite{BotSniffer} detects bots within the same botnet since their activities have spatial-temporal correlation and similarity. BotMiner \cite{BotMiner} performs cross cluster correlation to identify groups of compromised machines that shares similar botnet communication and activity patterns. Sekar et al. \cite{specificationids} proposed a specification-based technique to detect attacks as deviations from a normal model. All above researches focus on developing specifications of hosts and routers (rather than network) and building protocol state machine to detect anomalies. Such `profile' needs manual analysis and prior knowledge to construct, which VMP only relies on runtime execution traces.

\subsection{Virtual Machine Introspection}

Garfinkel \cite{vmi-based} proposed an architecture for intrusion detection using virtual machine introspection (VMI) technique. An IDS guest is installed with the monitored guest, and it implemented an OS interface library to interact with the virtual machine monitor (VMM) interface to obtain the VM state of the monitored guest (e.g., memory, register, and I/O devices) to further construct higher-level OS structures (e.g., process and virtual memory). They provide six sample security policies and monitor them with a modified VMware Workstation. The ReVirt \cite{ReVirt} targets on moving security logging mechanism into a virtual machine to provide better integrity. The security logging mechanism works as a loadable kernel module in the host operating system to inspect the interrupts for logging and replying. Chen \textit{et al}. \cite{better} also stated that secure logging and intrusion detection could benefit from the virtualized environment.

VMwatcher \cite{Out-of-the-Box} overcame the semantic gap challenge to reconstruct internal semantic views (e.g., files, processes, and kernel modules) of a VM. VMwatcher proposed a view comparison-based detection to detect anomalies. They corroborate an internal view (generated from inside the VM) with an external view (generated from outside the VM) of the same objects of interest and detect the existence of hidden malware. Lares \cite{Payne,Lares} is a framework that can control an application running in an untrusted guest VM by inserting protected hooks into the execution flow of a process to be monitored. These hooks transfer control to a security VM that checks the monitored application using VMI and security policies. Ether \cite{Ether} can even perform VMI to analysis malware's system calls via Intel's VT-x hardware virtualization extensions on Xen hypervisor. Nitro \cite{Nitro} realizes system call hooks in KVM. We anticipate that VMI research is a promising approach for monitoring and logging malware activities due to its isolation and transparent property.

Cuckoo Sandbox \cite{cuckoo} is a malware analysis system that takes the advantage of virtualization technique as well. The most difference between Cockoo and VMP system is that we do not need to install any additional agent inside the guest OS. Cuckoo's agent is an in-guest XMLRPC server that helps the host to retrieve guest information. On the contrary, VMP leverages VMI technique to retrieve the process list information, the address of hooked dynamic loaded library, etc.

The main difference of our proposed mechanism is that we focus on the high-level Windows API calls to describe the activities taken by the malware, and we tracking all the behaviors from the viewpoint of tainted analysis. Every file or registry that a process (and its spawned processes) access is monitored and traced. Our profiling mechanism stays outside the VM Guest as a part of QEMU plug-in to inspect the guest information.

\subsection{Behavior Analysis}

In 1996, Forrest \cite{sense,anomaly} introduced a concept -- self -- for anomaly detection by defining a normal process using a short-range correlation of its system calls. They create a database by using the sequences of system calls from the normal processes. Then, any unseen sequence indicates anomalies of the current testing process. Their experiment showed the longer consecutive calls are used to define 'self', the more accurate result it is to detect an anomaly. Their experiments demonstrated the result of 5, 6, and 11 consecutive system calls.

Some behavior analysis focus on malware clustering and similarity problem. Bailey et al. \cite{autoclass} pointed out inconsistencies in labeling by anti-malware vendors and present an automated classification system. They also pointed out that system call may be at a level that is too low for abstracting semantically meaningful information. Thus, they defined the behavior of malware in terms of non-transient state changes (e.g., spawned process name, modified registry keys, and modified file names) that the malware causes on the system. The use normalized compression distance (NCD) as the distance function to construct relationships between malware.

Karim et al. \cite{Permutations} generate phylogeny for Internet Worms by comparing permuted variants of programs. The result may help forensic analysts investigate new specimens, and assist in reconciling malware naming inconsistencies.

Bayer et al. \cite{Scalable} proposed a profile based on the operating system objects (e.g., file or registry key) and system calls on Anubis \cite{dynamic}, and they focused on developing novel clustering technique that scales well and produces precise results.

Schultz et al. \cite{Schultz,Gandotra} were the first to introduce the concept of data mining for detecting malware. They used three different static features for malware classification: Portable Executable (PE), strings and byte sequences. Then, Kolter et al. \cite{Kolter} used n-gram sequence detect malicious executables.

Kong et al. \cite{Kong} present a malware classification based on function call graph with a learning algorithm. Islam et al. \cite{Islam} used both static and dynamic features to classify malware. Leita et al. \cite{landscape} combined clustering techniques based on static and behavioral characteristics of the malware samples.	

As we seen above, several techniques are used in the past researches. However, we point out that some features in our proposed profile, such as high-level APIs, runtime parameters and return values, spawned child processes, are the essential to malware behavior profiling.

\section{Conclusions}
\label{sec:conclusion}

In this work, we propose the VMP scheme to generate malware profile and construct phylogenetic tree to group malware behaviors into families. We then discuss the importance of malware family and its identification. We also proposed the Pairwise Classification Score to evaluate the goodness of family classification of multiple anti-virus engines using different malware family naming scheme.

The VMI approach is engineered to record the API calls with parameters and return values as the malware profile. The VMP approach bridges the semantics between machine execution and program notation. The generated phylogenetic trees can appropriately identify malware families and clearly distinguish between benign programs and variants of a malware family. As there is no benchmark for malware family construction, we propose a voting-like intelligence based approach, called PCS, to compare the coherency of a detection engine with all others. The results show VMP performs better than most of other malware detection engines, and compatible with Text\_Mining and N-Gram\_N approach.

In addition, we construct a website (\url{http://140.112.107.39/index/}) for the proposed VMP system, which is built on the top of a distributed task queue system and multiple virtual machines running VMP to perform malware profiling. An uploaded executable will be profiled by VMP for 5 minutes and the corresponding profile can be downloaded on the website later.

It deserves further investigations to find out the set of API calls and resources substantial to distinguish different malware families. In addition, we only considered API calls for profile generation in this study, while other information, such as network communications and packet contents, could allow a better understanding of malware behavior. The inclusion of such information in malware profiling will be our next task. In this paper, we only present the VMP scheme and a possible approach that we do not investigate for best similarity/distance functions and machine learning approaches for malware family construction, neither do we fine tune on the thresholds used in VMP, which we believe there is much room for improvement.


%




\ifCLASSOPTIONcaptionsoff
  \newpage
\fi



%

%

\begin{IEEEbiography}[{\includegraphics[width=1in,height=1.25in,clip,keepaspectratio]{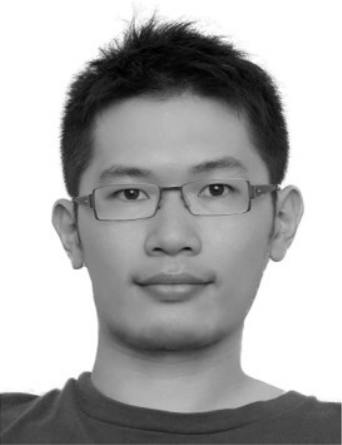}}]{Shun-Wen Hsiao}
received his B.S and Ph.D. degree from the Department of Information Management from National Taiwan University in 2004 and 2012, respectively. From 2006 to 2008, he participated in the iCAST collaborate research project with the CyLab of Carnegie Mellon University. Since July 2012, he joined Institute of Information Science, Academia Sinica, Taiwan and held postdoctoral research fellowship. In 2017, he joints the Department of Management Information Systems, National Chengchi University where he holds a position as an assistant professor. His research interests are in the area of computer networks, network security, virtualization technology, and FinTech.
\end{IEEEbiography}

\begin{IEEEbiography}[{\includegraphics[width=1in,height=1.25in,clip,keepaspectratio]{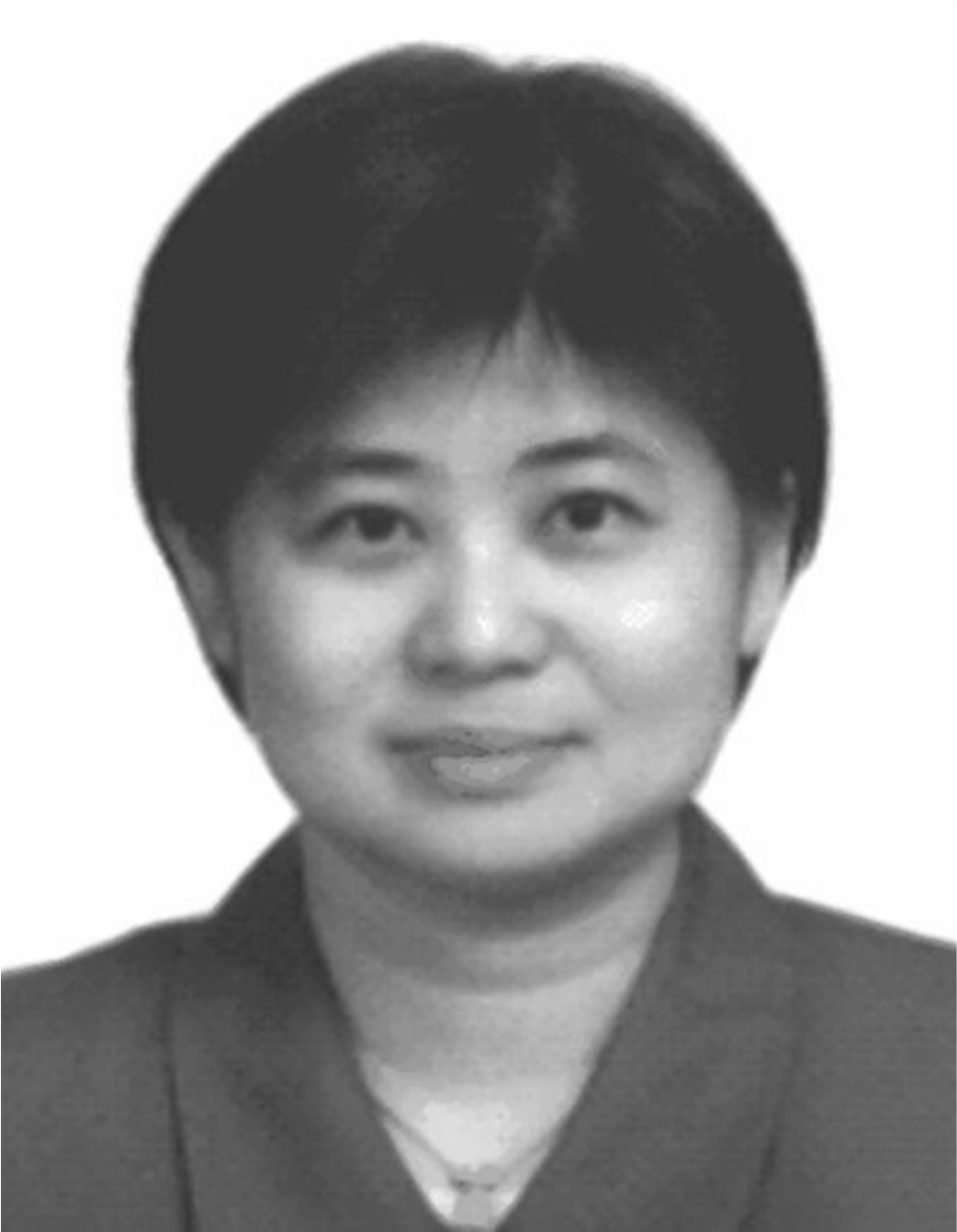}}]{Yeali S. Sun}
received her B.S. degree in Computer Science from National Taiwan University, and the M.S. and Ph.D. degrees in Computer Science from the University of California, Los Angeles (UCLA) in 1984 and 1988, respectively. From 1988 to 1993, she was with Bell Communications Research Inc. (Bellcore), where she was involved in the area of planning and architecture design of information networking, broadband networks, and network and system management. In 1993, she jointed the Department of Information Management, National Taiwan University where she holds a position as a professor now, and served as department head from 2006 to 2008. She was the director of the Compute and Information Networking Center of National Taiwan University from 2009 to 2014. From 1996 to 2002, she served in the TANet Technical Committee, Steering Committee of the National Broadband Experimental Network (NBEN) and Internet2, and IP Committee of TWNIC. Her research interests are in the areas of system and network security, quality of service (QoS), wireless mesh networks, multimedia content delivery, Internet pricing and network management, and performance modeling and evaluation.
\end{IEEEbiography}

\begin{IEEEbiography}[{\includegraphics[width=1in,height=1.25in,clip,keepaspectratio]{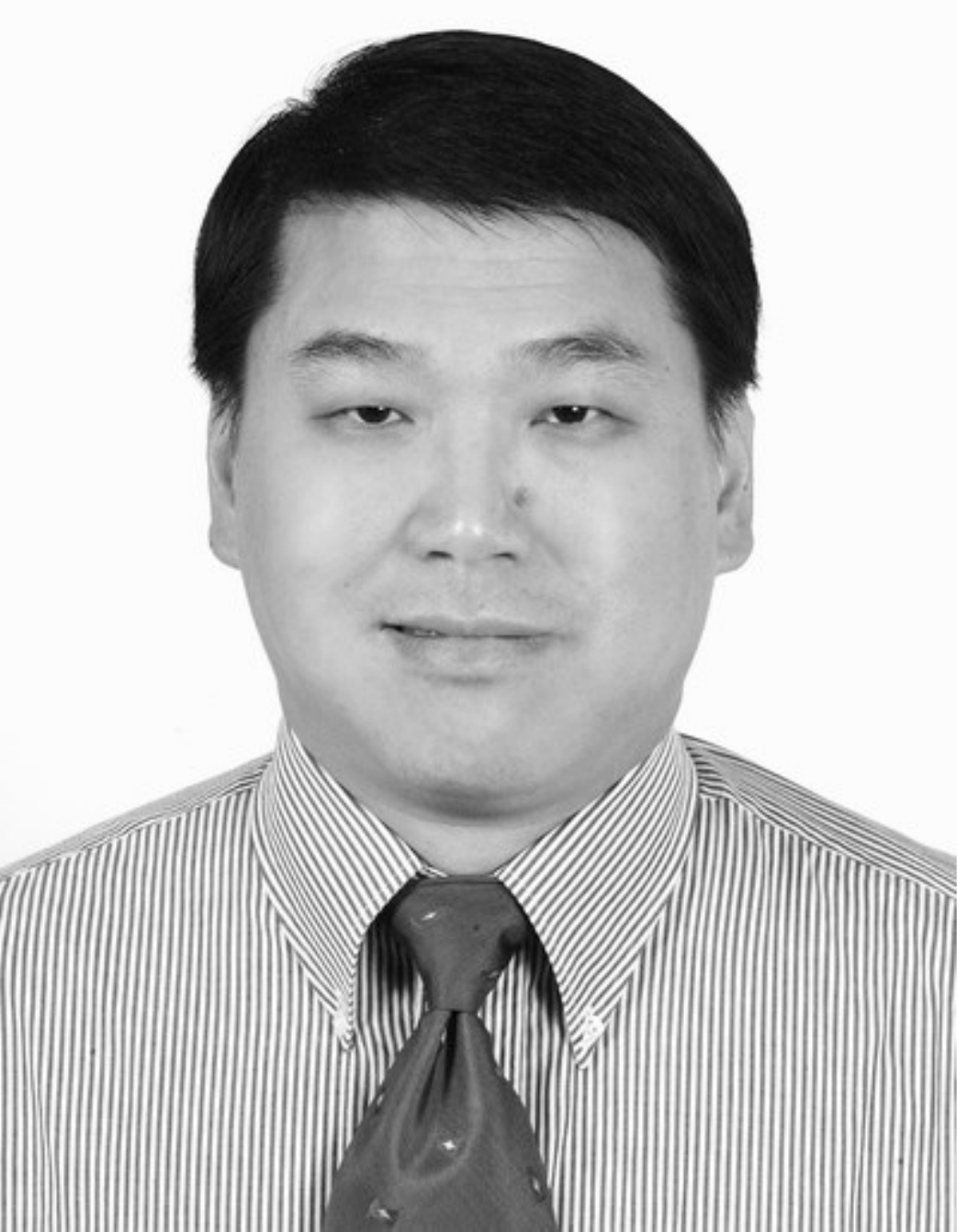}}]{Meng Chang Chen}
received the B.S. and M.S. degrees in Computer Science from National Chiao Tung University, Taiwan and the Ph.D. degree in Computer Science from the University of California, Los Angeles, in 1989. He joined AT\&T Bell Labs in 1989 as Member of Technical Staff and as technical leader of several projects in the area of data quality of distributed databases for mission critical systems. Since 1993 he has been with Institute of Information Science, Academia Sinica, Taiwan and assumed the responsibility of Deputy Director for 5 years. He is currently a Research Fellow. His current research interests include wireless network, network security, information retrieval, and data and knowledge engineering.
\end{IEEEbiography}





\vfill


\end{document}